\documentclass[aps,pra,reprint, amsmath, amssymb,superscriptaddress]{revtex4-2}
 
\usepackage{bm}
\usepackage[retainorgcmds]{IEEEtrantools}
\usepackage{graphicx}
\usepackage{mathrsfs}
\usepackage{amsmath}
\usepackage{amsfonts}
\usepackage{amssymb}
\usepackage{bbm}
\usepackage{color}
\usepackage{times,txfonts}
\usepackage{nicefrac}
\usepackage{ragged2e}
\usepackage{tikz}

\usepackage[colorlinks=true,linkcolor=blue,urlcolor=blue,citecolor=blue,pdfusetitle]{hyperref}

\usepackage{blindtext}

\newcommand{\tr}{\mathrm{tr}}
\newcommand{\trans}{^{\text{T}}}
\newcommand{\draz}{\text{+}} 
\newcommand{\rateMat}{\mathbb{W}}
\newcommand{\rateMatTL}{W}
\newcommand{\ident}{\mathbbm{1}} 

\newcommand{\Nb}{\bar{N}}
\newcommand{\toyxi}{\xi} 

\newcommand{\fall}{f^\text{(all)}}
\newcommand{\femp}{f^\text{(emp)}}

\newcommand{\Fiid}[1]{F_{#1}^\text{(iid)}}
\newcommand{\Fall}[1]{F_{#1}^\text{(all)}}
\newcommand{\Femp}[1]{F_{#1}^\text{(emp)}}
\newcommand{\Favg}[1]{F_{#1}^\text{(avg)}}

\newcommand{\sampMean}{\bar{X}}

\newcommand{\expect}[2][]{\def\tempa{#1}\def\tempb{} \text{E}\if\tempa\tempb #1 {_#1}\fi\left[#2\right]} 
\newcommand{\probof}[2][]{\def\tempa{#1}\def\tempb{} \text{P}\if\tempa\tempb #1 {_#1}\fi \left( #2 \right)}
\newcommand{\probgiven}[3][]{\def\tempa{#1}\def\tempb{} \text{P}\if\tempa\tempb #1 {_#1}\fi \left( #2 \middle\vert #3 \right)}

\newcommand{\ket}[1]{\left| #1 \right>}

\graphicspath{{figs/}}

\begin{document}

\title{Stochastic metrology and the empirical distribution}
\date{May 25, 2023}
\author{Joseph A. Smiga}
\email{joseph.smiga@rochester.edu}
\affiliation{Department of Physics and Astronomy, University of Rochester, Rochester, New York 14627, USA}
\author{Marco Radaelli}
\email{radaellm@tcd.ie}
\affiliation{School of Physics, Trinity College Dublin, Dublin 2, Ireland}
\author{Felix C. Binder}
\email{quantum@felix-binder.net}
\affiliation{School of Physics, Trinity College Dublin, Dublin 2, Ireland}
\email{quantum@felix-binder.net}
\author{Gabriel T. Landi}
\email{gabriel.landi@rochester.edu}
\affiliation{Department of Physics and Astronomy, University of Rochester, Rochester, New York 14627, USA}

\begin{abstract}

We study the problem of parameter estimation in time series stemming from general stochastic processes, where the outcomes may exhibit arbitrary temporal correlations.
In particular, we address the question of how much Fisher information is lost if the stochastic process is compressed into a single histogram, known as the empirical distribution.
As we show, the answer is non-trivial due to the correlations between outcomes. 
We derive practical formulas for the resulting Fisher information for various scenarios, from generic stationary processes to discrete-time Markov chains to continuous-time classical master equations. 
The results are illustrated with several examples. 

\end{abstract}

\maketitle{}

\section{Introduction}\label{sec:introduction}

\begin{figure*}
    \centering
    \includegraphics[width=\textwidth]{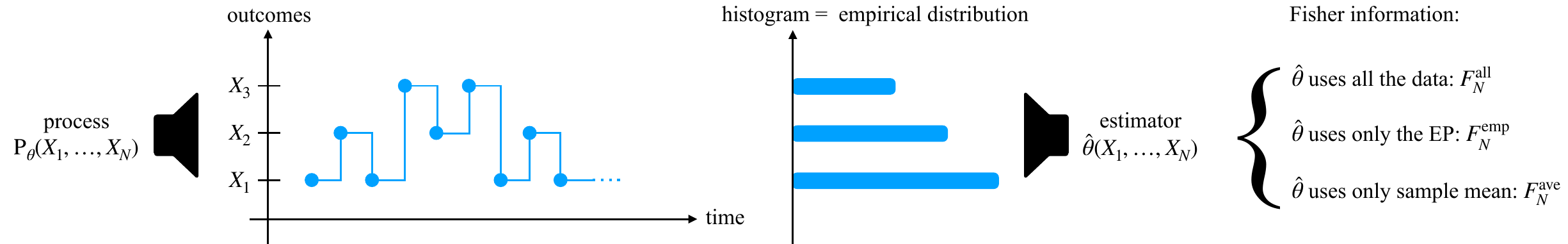}
    \caption{Basic paradigm in stochastic metrology. Data are generated from a stochastic process described by the probability $\probof[\theta]{X_1,\ldots,X_N}$ which depends on a unknown parameter $\theta$. This can be easily extended to the limit of continuous monitoring in time. A histogram of measurements can be derived, defining the empirical distribution [Eq.~\eqref{eq:empDist}], or compressed further in terms of the sample mean [Eq.~\eqref{eq:sample_mean}].
    The corresponding Fisher information for the derived quantities, which determines the ultimate precision of the estimation [Eq.~\eqref{eq:CR}], will depend on which aspects of the data the estimator $\hat{\theta}$ utilizes, and is bounded according to Eq.~\eqref{eq:3_fisher_inequalities}.
    }
    \label{fig:drawing}
\end{figure*}

In the standard scenario of parameter estimation~\cite{gammelmark_bayesian_2013, gammelmark_fisher_2014, cover_elements_2006, kay_fundamentals_1993, van_trees_detection_2013}, data are produced from a source that depends on an unknown parameter $\theta$. The aim is to estimate $\theta$ from the data.
An experimenter will repeat the experiment $N$ times, and collect  data points $\bm{X} = (X_1,\ldots,X_N)$ which can be visualized in a histogram (Fig.~\ref{fig:drawing}).
From this, one then builds an estimator $\hat{\theta}(\bm{X})$, which provides a guess as to the real value of  $\theta$. 
The precision of this estimation is fundamentally limited by the Cram\'er-Rao bound~\cite{rao_information_1992, nielsen_cramer-rao_2013, harald_cramer_mathematical_2016}
\begin{equation}\label{eq:CR}
    {\rm var}\Big(\hat{\theta}(\bm{X})\Big) \geqslant \frac{\Big[\partial_\theta \expect{\hat{\theta}(\bm{X})}\Big]^2}{F_N}\,,
\end{equation}
where $F_N$ is the Fisher Information (FI) contained in the distribution $\probof[\theta]{\bm{X}} = \probof[\theta]{X_1,\ldots,X_N}$: 
\begin{equation}\label{eq:fisher}
    F_N= \sum_{\bm{X}} \frac{\left[ \partial_\theta \probof[\theta]{\bm{X}}\right]^2}{\probof[\theta]{\bm{X}}}\,.
\end{equation}
The FI is a fundamental object in statistics and information theory, describing the ultimate precision achievable in estimating the parameter $\theta$ from the data at hand. 
The simplest scenario is when the outcomes are independent and identically distributed (iid) so that 
$\probof[\theta]{\bm{X}} = \probof[\theta]{X_1} \cdots \probof[\theta]{X_N}$.
The Fisher information in this case simplifies to being linear in $N$:
\begin{equation}\label{eq:F_iid}
    \Fiid{N} = N\sum_X \frac{\big[ \partial_\theta \probof[\theta]{X}\big]^2}{\probof[\theta]{X}}\,.
\end{equation}
In this paper, we consider a different paradigm, in which the data are not iid, but instead obtained as the outcomes of a stochastic process that exhibits temporal correlations between measurements. 
This represents the \emph{continuous monitoring} of a system, as it undergoes stochastic transitions between different states. 
Examples include Brownian motion or classical master equations~\cite{risken1989,coffey2004,kampen_stochastic_2007}, as well as continuous measurement of a quantum system in terms of quantum trajectories~\cite{gammelmark_bayesian_2013, gammelmark_fisher_2014, kiilerich_bayesian_2016, hasegawa_unifying_2023, hasegawa_thermodynamic_2023}.
Since each outcome now depends on the previous ones, the data are not iid and, as a consequence, the FI is no longer given by Eq.~\eqref{eq:F_iid}. 
Our goal is to determine what it is. 
As we will show, this crucially depends on what aspects of the dataset $\bm{X} = (X_1,\ldots,X_N)$ one uses in the estimator.

Maximum precision is achieved with estimators $\hat{\theta}(\bm{X})$ exploiting every possible aspect of the stochastic process, generally resulting in complicated functions.
For example, one might use estimators based on the likelihood of specific transitions chains $x \to y$ or $x \to y \to z$ and so on. 
The Fisher information of the full dataset $\Fall{N}$ provides a limit to the precision achievable from such an estimator.
Oftentimes, however, one might simply make a histogram of the outcomes.
This defines the \emph{empirical distribution}, which is illustrated in Fig.~\ref{fig:drawing}: suppose that each $X_i$ can take values within a certain alphabet $x \in \mathcal{A}$ (which may not be labeled by numbers). Then the empirical distribution (ED) is defined as 
\begin{equation}\label{eq:empDist}
    q_x = \frac{1}{N} \sum_{i=1}^N \delta_{X_i,x}\,.
\end{equation}
That is, it describes the proportion of observations for each possible outcome $x$. 
The Fisher Information $\Femp{N}$ associated with an estimator $\hat{\theta}(\bm{q})$ that uses only the empirical distribution can at best be as large as $\Fall{N}$ as one cannot gain additional information about $\theta$ when deriving the former from the latter.
Unlike the case of iid outcomes, in a correlated stochastic process the \emph{order} in which the data appear is of crucial importance. That is, while the ED contains information on how often states were measured, it does not contain information on the order of the measurements, which can be relevant when $\theta$ affects correlations between measurements. 
One can also analogously define empirical distributions counting higher order statistics, such as the number of transitions from $x\to y$, or from $x\to y \to z$ and so on. This extension of the empirical distribution to finite sequences potentially provides additional information, having Fisher information bounded (inclusively) between $\Femp{N}$ and $\Fall{N}$.

One could compress the data even more and consider the sample mean 
\begin{equation}\label{eq:sample_mean}
    \sampMean = \frac{J_{X_1}+\cdots+J_{X_N}}{N} = \sum_{x} J_x q_x\,,
\end{equation}
where $J_x$ is some mapping of the state labels $x$ to numeric values.
Since the sample mean can be constructed from the empirical distribution, its FI $\Favg{N}$ in turn cannot exceed $\Femp{N}$. 
A natural hierarchy arises then among these parameter estimation strategies: 
\begin{equation}\label{eq:3_fisher_inequalities}
    \Favg{N} \leqslant \Femp{N} \leqslant \Fall{N}\,. 
\end{equation}
These three quantities will generally differ from $\Fiid{N}$ [Eq.~\eqref{eq:F_iid}] computed on the single-symbol marginals, which could in principle be larger or smaller than $\Fall{N}$~\cite{radaelli_fisher_2022}.


\subsection{Summary of the main results}

In this paper we provide fundamental new results on how to compute the above-mentioned FI quantities and how to interpret them for a wide variety of stochastic processes. 
These results are relevant because building estimators exploiting the full information in the stochastic process may be very complicated when temporal correlations are present. 
The quantity $\Femp{N}$ provides the actual accessible information about the parameter if only the ED is taken into account. 

The paper starts in Sec.~\ref{sec:allFish} with a review of the Fisher information in the full dataset.
In Sec.~\ref{sec:empFish}, the Fisher information in the empirical distribution is calculated, followed by the calculation of the Fisher information in the sample mean in Sec.~\ref{sec:avgFish}.
Finally, in Sec.~\ref{sec:extension_ED} we extend the results for a higher order ED, accounting for transitions between states.
The results are illustrated through examples in Sec.~\ref{sec:examples} and conclusions are given in Sec~\ref{sec:conclusions}. 
A summary of our main results is shown in Table~\ref{tab:results_summary}, which covers the various Fisher information measures for different classes of stochastic processes, starting with very general stationary stochastic processes, then specializing to Markov chains and finally to continuous-time (Pauli) master equations.

\begin{table*}[ht]
    \centering
    \caption{Fisher information of a stationary stochastic process and its compression in terms of the empirical distribution and the sample mean. 
    The first column $\Fall{}$ is the full Fisher information in the process; 
    the second $\Femp{}$ is the Fisher information if we compress the full dataset in terms only of the empirical distribution, Eq.~\eqref{eq:empDist};
    and the third is the information contained in the sample mean, Eq.~\eqref{eq:sample_mean}.
    The results refer to a stochastic process with $N$ outcomes in the case of discrete-time, or a total time $\tau$ in the case of continuous time. 
    The first row, ``general process'' refers to a generic stationary stochastic process. It assumes finite Markov order-$M$ Markov for $\Fall{}$ and, for $\Femp{}$ and $\Favg{}$ that the statistics of the observed empirical distribution follow a normal distribution. The quantity $F_{M+1|1\ldots M}$ is the Fisher information given the previous $M$ measurements (see~\cite{radaelli_fisher_2022}). The matrix $\Psi$, defined in Eq.~\eqref{eq:Psi_matrix}, encodes the correlation between measurements. The transition matrix is $Q$, while the rate matrix is $\rateMat$. The steady state is $\boldsymbol{p}$ and $\mathbb{P}=\text{diag}\,\boldsymbol{p}$. The vector $\bm{J}$ contains numeric values for the states while $\mu$ is the expected sample mean using these numeric values. }
    \begin{tabular}{l lc lc lc}
    \hline\hline
    &$\Fall{}$ && $\Femp{}$ && $\Favg{}$ \\
    \hline
    General process & \eqref{eq:Fall_Markov_M} & $N F_{M+1|1\ldots M}$ & 
        \eqref{eq:Femp_general_stoch_process} & $N (\partial_\theta \bm{p})\trans \Big( \mathbb{P} + \Psi \mathbb{P} + \mathbb{P} \Psi\trans\Big)^{-1} (\partial_\theta \bm{p})$ & \eqref{eq:Favg_general_stoch_process} & $N \frac{(\partial_\theta \mu)^2}{\boldsymbol{J}\trans\left( \mathbb{P} + \Psi \mathbb{P} + \mathbb{P} \Psi\trans\right)\boldsymbol{J}}$\\
    Markov ($M=1$) & \eqref{eq:Fall_Markov_1} & $N\sum_y p_y \sum_x \frac{\big[\partial_\theta Q_{xy}\big]^2}{Q_{xy}}$ & 
        \eqref{eq:Femp_Markov_1} & $N (\partial_\theta \bm{p})\trans \Bigg( \mathbb{P} + Q(1-Q)^\draz\mathbb{P} + \mathbb{P} (1-Q\trans)^\draz Q\trans\Bigg)^{-1} (\partial_\theta \bm{p})$ & \eqref{eq:Favg_Markov_1} & $N \frac{(\partial_\theta \mu)^2}{\boldsymbol{J}\trans \left( \mathbb{P} + Q(1-Q)^\draz\mathbb{P} + \mathbb{P} (1-Q\trans)^\draz Q\trans \right) \boldsymbol{J}}$ \\
    Time-continuous &\eqref{eq:Fall_Markov_tcont} & $\tau \sum_y p_y \sum_{x\neq y} \frac{\big[ \partial_\theta \rateMat_{xy}\big]^2}{\rateMat_{xy}}$ & 
        \eqref{eq:Femp_master_no_DB} & $-\tau (\partial_\theta \bm{p})\trans \Bigg( \rateMat^\draz \mathbb{P} + \mathbb{P} (\rateMat^\draz)\trans\Bigg)^\draz (\partial_\theta \bm{p})$ & \eqref{eq:avgFish_cMark} & $-\tau \frac{(\partial_\theta \mu)^2}{ \boldsymbol{J}\trans \left( \rateMat^\draz \mathbb{P} + \mathbb{P} (\rateMat^\draz)\trans \right)^\draz \boldsymbol{J}}$ \\
    + detailed balance & & --- & 
        \eqref{eq:Femp_master_DB} & $- \frac{\tau}{2} (\partial_\theta \bm{p})\trans \mathbb{P}^{-1} \rateMat (\partial_\theta \bm{p})$ & \eqref{eq:Favg_master_DB} & $-\frac{\tau}{2} \frac{(\partial_\theta \mu)^2}{ \boldsymbol{J}\trans \left( \rateMat^\draz \mathbb{P} \right)^\draz \boldsymbol{J}}$ \\
    \hline\hline
    \end{tabular}
    \label{tab:results_summary}
\end{table*}

To motivate the following sections, we briefly present here the results for the particular case of master equations, in light of their importance in physics. Such systems exhibit correlations that propagate in time while still being simple enough that their dynamics can often be well understood.
Consider a system evolving according to the continuous time classical (Pauli) master equation~\cite{kampen_stochastic_2007} 
\begin{equation}\label{eq:master}
    \frac{d p_x}{dt} = \sum_{y} \left( \rateMatTL_{xy} p_y - \rateMatTL_{yx} p_x\right) := \sum_{y} \rateMat_{xy} p_y\,,
\end{equation}
where $\rateMatTL_{xy}$ is the transition rate from $y\to x$ and $\rateMat$ is a matrix with off-diagonal elements $W_{xy}$ and diagonal elements given by the escape rates $\rateMat_{xx}=-\sum_y \rateMatTL_{xy}$. 
The parameter $\theta$ is encoded in the transition rates $\rateMatTL_{xy}$.

Here and henceforth we denote the steady-state of Eq.~\eqref{eq:master}  by $\bm{p}$ (i.e., $\rateMat \bm{p} = 0$). Throughout, it will be assumed that this solution exists (i.e., the process is aperiodic) and is unique (i.e., the model is irreducible).
The FI for the full stochastic trajectory of duration $\tau$ is given by (Sec.~\ref{sec:allFish_cont}):
\begin{equation}\label{eq:fall}
    \Fall{\tau} = \tau \sum_y p_y \sum_{x\neq y} \frac{\big[ \partial_\theta \rateMatTL_{xy}\big]^2}{\rateMatTL_{xy}}\,,
\end{equation}
Our main contribution is the FI of the empirical distribution. In the case of systems satisfying detailed balance, the formula becomes particularly simple and reads
\begin{equation}\label{eq:femp_DB}
    \Femp{\tau} = -\frac{\tau }{2}\sum_{x,y} (\partial_\theta p_x) \frac{W_{xy}}{p_x} (\partial_\theta p_y)\,.
\end{equation}
This result shows noteworthy similarities with the the iid case in Eq.~\eqref{eq:F_iid}. The fundamental difference, however, is now the role played by the transition matrix.
Although not evident, Eqs.~\eqref{eq:fall} and~\eqref{eq:femp_DB} are bounded by the inequality~\eqref{eq:3_fisher_inequalities}. 
The difference between them, therefore, summarizes the amount of information that is not present in the steady-state distribution $\bm{p}$, but is encoded in the transition rates $\rateMatTL_{xy}$.

\section{Fisher information of stationary stochastic processes}\label{sec:allFish}

We start by computing the full FI $\Fall{N}$ for a stationary stochastic process $X_1,\ldots,X_N$  described by a distribution $\probof{X_1=x_1,\ldots,X_N=x_n} \equiv \probof{x_1,\ldots,x_N}$ encoding the parameter $\theta$ we wish to estimate; i.e., $\probof{\boldsymbol{X}} = \probof[\theta]{\boldsymbol{X}}$.
For the sake of simplicity, we will consider single parameter estimation in what follows. However, the extension to the multi-parameter case is straightforward.
In a stationary processes, for any block of size $r< N$, 
\begin{equation}\label{eq:stationaryProc}
    \probof{X_1,\ldots,X_r} = \probof{X_{1+k},\ldots,X_{r+k}}\,.
\end{equation}
We assume each $X_i$ can take values within a certain alphabet $\mathcal{A}$, and we denote the reduced distribution of a single outcome as $p_x = \probof{X_i = x}$ (which is independent of $i$ for stationary processes). 

\subsection{General Markov process}

We first consider the case of a process with a finite Markov order $M$: a process with $M=0$ has iid outcomes, a process with $M=1$ (a Markov chain) is decomposed as 
\begin{equation}\label{eq:Markov1}
    \probof{x_1,\ldots,x_N} = \probgiven{x_N}{x_{N-1}} \cdots \probgiven{x_2}{x_1} \probof{x_1}\,,
\end{equation}
a process with $M=2$ is decomposed as 
\begin{equation}
    \probof{x_1,\ldots,x_N} = \probgiven{x_N}{x_{N-1},x_{N-2}} \cdots \probgiven{x_3}{x_2,x_1} \probof{x_2,x_1}\,,
\end{equation}
and so on. 
Further generalizing such processes, one can consider a Hidden Markov Model (HMM)~\cite{baum_statistical_1966, ephraim_hidden_2002}, which builds on a finite-order Markov process, but whose outcomes follow a random distribution parameterized by the state of the underlying Markov process. HMM can have infinite Markov order because information in recent outcomes of the underlying process is lost; meaning that subsequent states in the HMM cannot be perfectly modeled based only on finitely many past states. Even in the case of infinite Markov order, a finite HMM representation may suffice. Memory-minimal HMMs for stationary processes are known as $\varepsilon$-machines~\cite{Crutchfield1989}.
In practice, infinite Markov order cannot be determined from output data alone. However, one can often assign a sufficiently large, but finite, effective Markov order $M_{\rm eff}$, beyond which correlations can be neglected. 

The full Fisher information of $\probof{x_1,\ldots,x_N}$ is given by the usual definition 
\begin{equation}\label{Fisher_N_general}
    \Fall{N} = \sum_{x_1,\ldots,x_N} \frac{\big[ \partial_\theta \probof[\theta]{x_1,\ldots,x_N}\big]^2}{\probof[\theta]{x_1,\ldots,x_N}}\,.
\end{equation}
For finite Markov order $M< N$, we may use the result recently derived in Ref.~\cite{radaelli_fisher_2022}: 
\begin{equation}\label{eq:Marco}
    \Fall{N} = F_M + (N-M) F_{M+1|1\ldots M}\,.
\end{equation}
The first term is the Fisher information of $\probof{x_1,\ldots,x_M}$, while the second is a conditional Fisher information, given by 
\begin{equation}
    F_{M+1|1\ldots M} = \sum_{x_1,\ldots,x_M} \probof{x_1,\ldots,x_M} \sum_{x_{M+1}} \frac{\Big[\partial_\theta \probgiven{x_{M+1}}{x_1,\ldots,x_M}\Big]^2}{\probgiven{x_{M+1}}{x_1,\ldots,x_M}}\,.
\end{equation}
Equation~\eqref{eq:Marco} shows that for $N\gg M$ the Fisher information will asymptotically behave as 
\begin{equation}\label{eq:Fall_Markov_M}
    \Fall{N} \simeq N F_{M+1|1\ldots M}\,. 
\end{equation}
This linear scaling with $N$ stems solely from the assumption of exponential decay of correlations, and allows us to define a \emph{Fisher information rate}
\begin{equation}
    \fall = \frac{\Fall{N}}{N} \simeq F_{M+1|1\ldots M}\,.
\end{equation}
One may interpret $\fall$ as the rate of information obtained in each outcome of the experiment. 
A high information rate means that the estimation precision will increase faster with the number of data points $N$. 
This type of linear scaling appears often in the literature, for example in the case of continuously measured quantum systems~\cite{gammelmark_fisher_2014}.

\subsection{Discrete- and continuous-time Markov processes}\label{sec:allFish_cont}

A system with Markov order $M=1$ [Eq.~\eqref{eq:Markov1}] is fully  characterized by a transition matrix $Q_{xy} := \probgiven{x}{y}$. The stationary distribution, $p_x$, is the solution of $\bm{p} = Q \bm{p}$. 
In this case Eq.~\eqref{eq:Fall_Markov_M} simplifies to
\begin{equation}\label{eq:Fall_Markov_1}
    \Fall{N} = N\sum_y p_y \sum_x \frac{\big[\partial_\theta Q_{xy}\big]^2}{Q_{xy}}\,.
\end{equation}
We can specialize this to continuous-time cases where the system evolves according to the master equation~\eqref{eq:master}.
This can be viewed as a discrete-time process with transition matrix $Q = 1 + \rateMat dt$, where $dt$ is an infinitesimal time step. 
For $x\neq y$ we then have 
\begin{equation}\label{eq:transMat_off_cont_limit}
    \frac{\big[\partial_\theta Q_{xy}\big]^2}{Q_{xy}} = dt \frac{\big[\partial_\theta W_{xy}\big]^2}{W_{xy}}\,,
\end{equation}
while for $x=y$ we have 
\begin{equation}\label{eq:transMat_on_cont_limit}
    \frac{\big[\partial_\theta Q_{xx}\big]^2}{Q_{xx}} = dt^2 (\partial_\theta \rateMat_{xx})^2\,.
\end{equation}
This contribution is of order $dt^2$, however, hence negligible compared to~\eqref{eq:transMat_off_cont_limit}.
Equation~\eqref{eq:Fall_Markov_1} therefore reduces to
\begin{equation}\label{eq:Fall_Markov_tcont}
    \Fall{\tau} = \tau \sum_y p_y \sum_{x\neq y} \frac{\big[ \partial_\theta \rateMatTL_{xy}\big]^2}{\rateMatTL_{xy}}\,,
\end{equation}
which is precisely Eq.~\eqref{eq:fall}, with $\tau = N dt$. 
This expression has appeared in the literature before, e.g. in~\cite{vo_unified_2022}, although we are unaware of any papers discussing its derivation and consequences in any detail.

\section{Fisher information of the Empirical distribution}\label{sec:empFish}

We now turn to the ED, as defined by Eq.~\eqref{eq:empDist}. It records the proportion of instances in which outcomes $X_i$ take each of the variates $x\in\mathcal{A}$ of the alphabet. 
In addition to being stationary, we assume the process is ergodic~\cite{kampen_stochastic_2007}, and hence has a unique stationary solution. 

\subsection{General process}

Given a single chain of $N$ outcomes $\left( X_1,X_2,\ldots,X_N \right)$ the ED~\eqref{eq:empDist} is a vector of random variables $\boldsymbol{q}$, with 
$\boldsymbol{q}\in[0,1]^d$ where $d=\lvert\mathcal{A}\rvert$ is the (possibly infinite) cardinality of $\mathcal{A}$, and with the additional restriction that $\sum_x q_x = 1$. 
We then consider the statistics of measuring $\boldsymbol{q}$.
The ED is an unbiased estimator of the steady-state (single-outcome) distribution $p_x = \probof{X_i = x}$:
\begin{align}
    \expect{q_x} &= \sum_{x_1,\ldots,x_N} \probof{x_1,\ldots,x_N} \frac{1}{N}\sum_{i=1}^N \delta_{x_i,x}  
    = p_x\,. \label{eq:empDist_mean}
\end{align}
Here $\expect{\bullet}$ represents the average over multiple realizations of the same experiment. 
Since $q_x$ is obtained from a finite number $N$ of outcomes in a single experiment, it will in general fluctuate. 
To compute the corresponding Fisher information, we will assume that for large $N$ the random vector $\boldsymbol{q}$ is approximately distributed as a multivariate Gaussian. 
While we expect this to be true for most stationary processes, we are unaware of any theorems explicitly proving it
\footnote{
We have also not been able to come up with an example of a stationary process that violates this hypothesis. For an example of a \emph{non}-stationary process which violates the hypothesis, 
consider a binary alphabet $\mathcal{A}=\{0,1\}$ in which the probability of measuring 1 increases the more 1's measured before; causing a type of ``symmetry breaking'' towards one of the states. The distribution of $\boldsymbol{q}$ would be bimodal, with the two peaks corresponding to the case in which mostly 0's and mostly 1's are measured.
}.

We may define a covariance matrix $\Sigma_{xy} = {\rm Cov}(q_x,q_y)$, of dimensions $d\times d$. 
The covariance will depend on the conditional probability of two outcomes,
\begin{equation}
p_{x\gets y}(k) = \probgiven{X_{i+k}=x}{X_{i}=y}\,,
\end{equation}
where the right-hand-side is independent of $i$ for a stationary process. For Markov processes, $p_{x\gets y}(k)$ is fully described by a product of transition matrices.
As shown in Appendix~\ref{app:covariance_ED}, the covariance matrix of $\boldsymbol{q}$ can be written as 
\begin{equation}\label{eq:ED_covariance_matrix}
    \Sigma_{xy} = \frac{1}{N} \left(p_x (\delta_{xy}- p_y) + \Psi_{xy} p_y + \Psi_{yx} p_x
    \right)\,,
\end{equation}
where 
\begin{equation}\label{eq:Psi_matrix}
    \Psi_{xy} = \sum_{k=1}^{N-1} \left(1-\frac{k}{N}\right) \Big[p_{x\gets y}(k) - p_x\Big]\,.
\end{equation}
The matrix $\Psi$ encodes correlations between measurements, with $\Psi\equiv 0$ for iid processes.

With this at hand, we show in  Appendices~\ref{app:F_emp_computation} and~\ref{app:proof_inverse} our main result; namely that the Fisher information contained in the ED reads
\begin{equation}\label{eq:Femp_general_stoch_process}
    \Femp{N} = N (\partial_\theta \bm{p})\trans \Big( \mathbb{P} + \Psi \mathbb{P} + \mathbb{P} \Psi\trans\Big)^{-1} (\partial_\theta \bm{p})\,,
\end{equation}
where $\mathbb{P} = {\rm diag}(p_x)$ is a matrix with the steady-state distribution in its diagonals. 
As a sanity check, for iid outcomes $\Psi=0$ and we recover $\Fiid{N}$ in Eq.~\eqref{eq:F_iid}.
Therefore Eq.~\eqref{eq:Femp_general_stoch_process} shows how the correlations in the stochastic process, characterized by the matrix $\Psi$, modify the rate at which we acquire information about a process. 
Compared to Eq.~\eqref{eq:Fall_Markov_M}, it is always true that $\Femp{N} \leqslant \Fall{N}$, but in general there are no bounds relating $\Femp{N}$ and $\Fiid{N}$, so correlations can be both beneficial or deleterious for the acquisition of information~\cite{radaelli_fisher_2022}.

\subsection{Discrete-time Markov processes}

As before, we now specialize this to the case of Markov order $M=1$, where the system is characterized by the transition matrix $Q_{xy}$.
The transition probabilities $p_{x\gets y}(k)$ appearing in Eq.~\eqref{eq:Psi_matrix} are $p_{x\gets y}(k) = (Q^k)_{xy}$, and
\begin{equation}
    \Psi = \sum_{k=1}^{N-1} \left(1-\frac{k}{N}\right) \left[ Q^k - \boldsymbol{p}\boldsymbol{u}\trans \right]\,,
\end{equation}
where $\bm{u} = (1,\ldots,1)$ is a column vector with all entries equal to 1. 
To carry out the remaining sum, we assume that $Q$ is diagonalizable as~\footnote{When $Q$ is not diagonalizable a similar construction can also be made using Jordan blocks, leading to the same final result.}
\begin{equation}
    Q = \bm{p}\bm{u}\trans + \sum_j \lambda_j \bm{x}_j \bm{y}_j\trans\,, 
\end{equation}
where $|\lambda_j|<1$ are the eigenvalues and $\bm{x}_j, \bm{y}_j$ the corresponding right and left eigenvectors. 
Observe that $\left(\boldsymbol{p}\boldsymbol{u}\trans\right)^k = \boldsymbol{p}\boldsymbol{u}\trans$ by the normalization of $\boldsymbol{p}$ and $\boldsymbol{u}\trans\boldsymbol{x_j} = \boldsymbol{y_j}\trans\boldsymbol{p}$ by the decomposition, so only the $\lambda_j$ terms in $Q^k - \boldsymbol{p}\boldsymbol{u}\trans$ remain in the sum.
For $\lvert \alpha \rvert < 1$,
\begin{align*}
    \sum_{k=1}^{N-1} \left(1-\frac{k}{N}\right) \alpha^k &= \frac{\alpha}{1-\alpha} \left(1-\frac{1}{N}\frac{1-\alpha^N}{1-\alpha}\right) \\
    &\overset{N\gg 1}{\approx} \frac{\alpha}{1-\alpha}\,.
\end{align*}
It then follows that 
\begin{equation}\label{Psi_Markov_1}
    \Psi = \sum_j \frac{\lambda_j}{1-\lambda_j} \bm{x}_j \bm{y}_j\trans := Q(1-Q)^\draz\,.
\end{equation}
Here $(1-Q)^\draz$ is the Drazin pseudo-inverse of $1-Q$, which corresponds to inverting all eigenvalues which are not zero.  
For more details on the Drazin inverse, see Appendix~K in Ref.~\cite{kewming_diverging_2022}. 
Finally, plugging Eq.~\eqref{Psi_Markov_1} into Eq.~\eqref{eq:Femp_general_stoch_process} we obtain 
\begin{equation}\label{eq:Femp_Markov_1}
    \Femp{N} = N (\partial_\theta \bm{p})\trans \Bigg( \mathbb{P} + Q(1-Q)^\draz\mathbb{P} + \mathbb{P} (1-Q\trans)^\draz Q\trans\Bigg)^{-1} (\partial_\theta \bm{p})\,,
\end{equation}
which is the final form of $\Femp{N}$ for Markov order-1 processes. 
The expression is formally equivalent to~\eqref{eq:Femp_general_stoch_process}, but with the additional structure~\eqref{Psi_Markov_1} for the matrix $\Psi$.

\subsection{Continuous-time Markov processes}

Lastly, we consider the continuous-time case with $Q = 1 + \rateMat dt$. 
For large-$N$,
\begin{equation}
    Q(1-Q)^\draz \simeq - \frac{1}{dt}  \rateMat^\draz\,.
\end{equation}
Notice that $\rateMat$ is not invertible, so the Drazin inverse is needed once again. 
In the limit of small $dt$ Eq.~\eqref{eq:Femp_Markov_1} then reduces to 
\begin{equation}\label{eq:Femp_master_no_DB}
    \Femp{\tau} = -\tau (\partial_\theta \bm{p})\trans \left( \rateMat^\draz \mathbb{P} + \mathbb{P} (\rateMat^\draz)\trans\right)^\draz (\partial_\theta \bm{p})\,,
\end{equation}
where, again, $\tau = N dt$. 
This formula simplifies in the particular case where the system satisfies detailed balance: 
\begin{equation}\label{eq:detailed_balance}
    W_{xy}p_y = W_{yx} p_x \qquad \text{or} \qquad \rateMat\mathbb{P} = \mathbb{P} \rateMat\trans\,. 
\end{equation}
This is a common condition whenever the underlying system is described by a Hamiltonian that is an even function of momentum, and the observable states are also even functions of momentum~\cite{kampen_stochastic_2007}.
Taking the Drazin inverse of both sides shows that $\mathbb{P}(\rateMat^\draz)\trans = \rateMat^\draz \mathbb{P}$.
Finally, we use the fact that when acting between  vectors which sum to zero (see Eq.~\eqref{eq:drazOfProd} in Appendix~\ref{app:drazin}), as is the case with $\partial_\theta \bm{p}$ in Eq.~\eqref{eq:Femp_master_no_DB}, it follows that 
$(\rateMat^\draz \mathbb{P})^\draz = \mathbb{P}^{-1} \rateMat$. 
Hence Eq.~\eqref{eq:Femp_master_no_DB} reduces to 
\begin{equation}\label{eq:Femp_master_DB}
    \Femp{\tau} = - \frac{\tau}{2} (\partial_\theta \bm{p})\trans \mathbb{P}^{-1} \rateMat (\partial_\theta \bm{p})\,,
\end{equation}
which is Eq.~\eqref{eq:femp_DB}, written in matrix form. 
When detailed balance fails to hold, one must instead use Eq.~\eqref{eq:Femp_master_no_DB}.

\section{Fisher information of the sample mean}\label{sec:avgFish}

The sample mean, Eq.~\eqref{eq:sample_mean}, corresponds to a simple metrological protocol: given a stochastic process one simply computes the average of the quantity in question and uses that to build an estimator. 
Over multiple experiments, the sample mean will yield the correct average
\begin{equation}
    \mu := \expect{\bar{X}} = \sum_x J_x p_x = \bm{J}\trans \bm{p}\,,
\end{equation}
wherein $\boldsymbol{J}$ contains numeric labels for the corresponding states.
The variance of the sample mean for a stationary stochastic process follows from propagation-of-uncertainty for a linear transformation,
\begin{equation}
    \sigma_{\bar{X}}^2 = \bm{J}\trans \Sigma \bm{J}\,,
\end{equation}
where $\Sigma$ is the covariance matrix of the empirical distribution, Eq.~\eqref{eq:ED_covariance_matrix}. 

By an argument similar to that used for the ED, in the limit of large $N$ the sample mean will be approximately Gaussian so that we may once again employ Eq.~\eqref{F_Gaussian} to calculate the Fisher information with the uncertainty derived from the empirical distribution.
For the general process,
\begin{equation}\label{eq:Favg_general_stoch_process}
    \Favg{N} = N \frac{(\partial_\theta \mu)^2}{\boldsymbol{J}\trans\left( \mathbb{P} + \Psi \mathbb{P} + \mathbb{P} \Psi\trans\right)\boldsymbol{J}}\,.
\end{equation}
Restricting to order-1 Markov processes,
\begin{equation}\label{eq:Favg_Markov_1}
    \Favg{N} = N \frac{(\partial_\theta \mu)^2}{\boldsymbol{J}\trans \left( \mathbb{P} + Q(1-Q)^\draz\mathbb{P} + \mathbb{P} (1-Q\trans)^\draz Q\trans \right) \boldsymbol{J}}\,.
\end{equation}
For master equations,
\begin{equation}\label{eq:avgFish_cMark}
    \Favg{\tau} = - \tau \frac{(\partial_\theta \mu)^2}{\bm{J}\trans\big[\rateMat^{-1} \mathbb{P} + \mathbb{P} (\rateMat^{-1})\trans\big] \bm{J}}\,,
\end{equation}
which can be simplified when detailed balance is satisfied, similar to Eq.~\eqref{eq:Femp_master_DB},
\begin{equation}\label{eq:Favg_master_DB}
    \Favg{\tau} = -\frac{\tau}{2} \frac{(\partial_\theta \mu)^2}{ \boldsymbol{J}\trans \left( \rateMat^\draz \mathbb{P} \right)^\draz \boldsymbol{J}}\,.
\end{equation}

\section{Empirical distribution of a finite sequence}\label{sec:extension_ED}

The results of this paper can directly be extended to consider the empirical distribution describing how often a given finite sequence of measurements appears in a process. 
That is, for a length-$L$ sequence $\boldsymbol{x}=\{x_1,\ldots,x_L\}$, one can define the empirical distribution
\begin{equation}
    q(\boldsymbol{x}) = \frac{1}{N+1-L} \sum_{k=1}^{N+1-L} \delta_{x_1 X_{k}} \cdots \delta_{x_L X_{k+L-1}}\,.
\end{equation}
The mean value of $q(\boldsymbol{x})$ is
\begin{equation}
    \expect{q(\boldsymbol{x})} = \probof{x_1,\ldots,x_L} := \probof{\boldsymbol{x}}\,,
\end{equation}
as one may expect. 
It therefore provides a unbiased estimator of a length-$L$ sequence distribution.
For $L=1$ we recover Eq.~\eqref{eq:empDist}.
For $L=2$ we would be sampling the joint probabilities $P(x_1,x_2)$ and so on.

We continue to interpret $q(\boldsymbol{x})$ as a vector, but now with dimension $d^L$. 
To compute the corresponding Fisher information we will assume, as before, that in the large $N$ limit the vector $q(\boldsymbol{x})$ is jointly Gaussian. 
Some care must be taken when $L>1$, however, as ergodicity is insufficient to guarantee that every sequence of measurement outcomes may occur in a given process.
This can be avoided if we only consider entries of $q(\boldsymbol{x})$ over those states for which the transition probability is not identically zero. 
To find the covariance of $q(\boldsymbol{x})$, one first extends the conditional probability from before,
\begin{equation}\label{eq:procCorr_seq}
    p_{\boldsymbol{x}\gets\boldsymbol{y}}(k) = \probgiven{X_{k+1}=x_1,\ldots,X_{k+L}=x_L}{X_1=y_1,\ldots,X_{L}=y_L}\,.
\end{equation}
Observe that, for $k<L$, there will be some overlap between $\boldsymbol{x}$ and $\boldsymbol{y}$, thus such terms would include the delta functions
\footnote{For example, for $L=3$ and $k=1$, Eq.~\eqref{eq:procCorr_seq} describes the probability that $X_2=x_1$, $X_3=x_2$, and $X_4=x_3$ given that $X_1=y_1$, $X_2=y_2$, and $X_3=y_3$. In order for this to be true, one must demand $x_1=y_2$ and $x_2=y_3$ for the probability to be non-zero. Thus $p_{\boldsymbol{x}\gets\boldsymbol{y}}(1) = \delta_{x_1y_2}\delta_{x_2y_3}\probgiven{X_4=x_3}{X_1=y_1,X_2=x_1,X_3=x_2}$.}
$\delta_{y_{L+1-k}x_1}\cdots\delta_{y_{L}x_{L-k}}$. 
Then, as in Eq~\eqref{eq:Psi_matrix}, we define the following matrix to encode the correlation,
\begin{equation}
    \Psi_{\boldsymbol{x}\boldsymbol{y}} = \sum_{k=1}^{N-L} \left(1-\frac{k}{N+1-L}\right) \left[ p_{\boldsymbol{x}\gets \boldsymbol{y}}(k)-\probof{\boldsymbol{x}} \right]\,.
\end{equation}
Finally, the covariance $\Sigma_{\boldsymbol{x}\boldsymbol{y}}=\text{Cov}\left( q(\boldsymbol{x}),q(\boldsymbol{y}) \right)$ is
\begin{align}
    \Sigma_{\boldsymbol{x}\boldsymbol{y}} &= \frac{1}{N+1-L} \left[ \probof{\boldsymbol{x}} \left( \delta_{\boldsymbol{x}\boldsymbol{y}} - \probof{\boldsymbol{x}} \right) + \right.\nonumber\\
    &\quad\qquad\qquad\qquad\left. \Psi_{\boldsymbol{x}\boldsymbol{y}}\probof{\boldsymbol{y}} + \probof{\boldsymbol{x}}\Psi_{\boldsymbol{y}\boldsymbol{x}} \right]\,.
\end{align}
Defining the sequence-indexed vector and matrix,
\begin{equation}
    p_{\boldsymbol{x}} = \probof{\boldsymbol{x}} \quad\text{and}\quad \mathbb{P}  =\text{diag}\left( \boldsymbol{p} \right)\,.
\end{equation}
the Fisher information then has the same form as Eq.~\eqref{eq:Femp_general_stoch_process}:
\begin{equation}\label{eq:Fseq}
    F^{\text{(emp,L)}}_{N} = N (\partial_\theta \bm{p})\trans \left( \mathbb{P} + \Psi \mathbb{P} + \mathbb{P} \Psi\trans\right)^{-1} (\partial_\theta \bm{p})\,.
\end{equation} 
Our results therefore readily generalize to the ED of arbitrary length-$L$ sequences.


\section{Examples}\label{sec:examples}

\begin{figure*}[ht]
    \centering
    \includegraphics{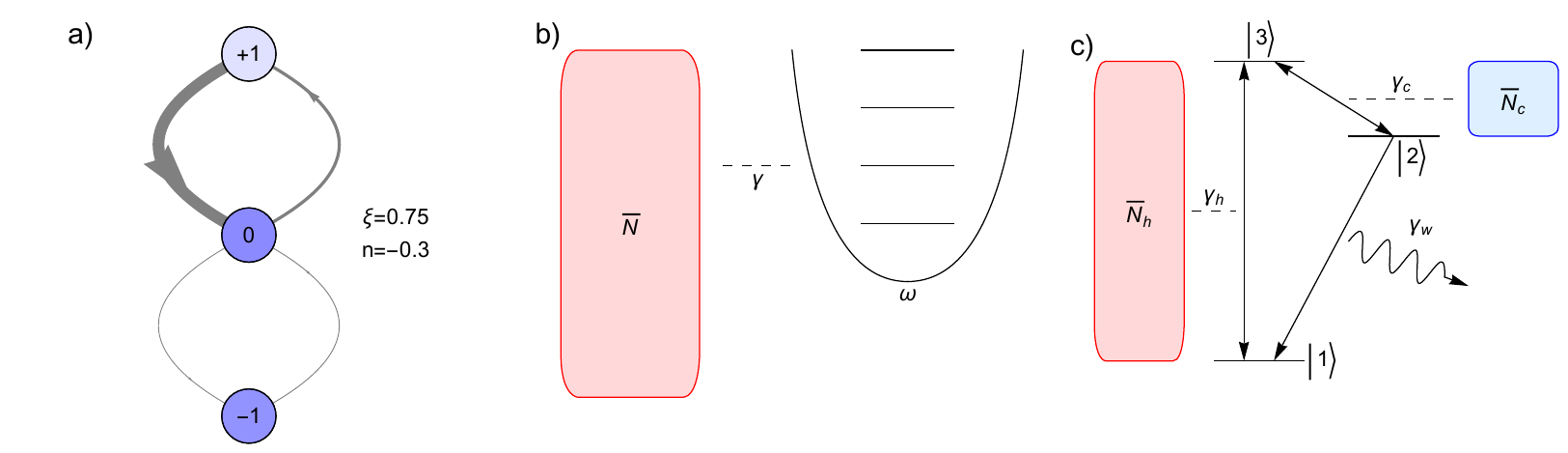}
    \caption{Diagrams of models considered. (a) The three-level toy model. Using $\beta=0.75$ and $n=-0.3$, the circles representing the states are darkened according to the steady-state solution. The thickness of the arrows between the states represents their respective transition rates. (b) A harmonic oscillator coupled to a thermal bath. (c) A three-level maser driven by a hot and cold thermal bath.}
    \label{fig:ex_setups}
\end{figure*}

We now illustrate our results with several examples. 
In Secs.~\ref{sec:examples_2level}-\ref{sec:examples_maser} we consider examples stemming from continuous time master equations. 
Then, in Sec.~\ref{sec:examples_spin_chain} we discuss an example of a discrete-step stochastic process.

\subsection{Two-level systems}
\label{sec:examples_2level}

The simplest example is a system containing two states in the alphabet and evolving according to a master equation with transition matrix 
\begin{equation}
    \mathbb{W} = \left[\begin{matrix}
        -a_\theta & b_\theta \\
        a_\theta & -b_\theta
    \end{matrix}\right]\,,
\end{equation}
with non-negative transition rates $a_\theta$ and $b_\theta$ that encode the parameter $\theta$ in question.
The functional dependence of the encoding is left arbitrary. 
For simplicity, we omit the $\theta$ dependence below, and write them only as $a$ and $b$. 
The steady-state is $\bm{p} = \left(\frac{b}{a+b}, \frac{a}{a+b}\right)$ and this system always satisfies detailed balance [Eq.~\eqref{eq:detailed_balance}].

The total Fisher information is given by Eq.~\eqref{eq:Fall_Markov_tcont} and reads 
\begin{equation}
    \Fall{\tau} = \tau \frac{\left( b\partial_\theta a\right)^2 + \left( a \partial_\theta b \right)^2}{a b (a+b)}\,.
\end{equation}
Conversely, the Fisher information of the ED [Eq.~\eqref{eq:Femp_master_DB}] is given by 
\begin{equation}
    \Femp{\tau} = \frac{\big( b \partial_\theta a - a \partial_\theta b\big)^2}{2a b (a+b)}\,.
\end{equation}
The two are only equal in the special case where $a \partial_\theta b = -b \partial_\theta a$. 
Otherwise, we have
\begin{equation}
    \Fall{\tau} = 2\Femp{\tau} + 2 \tau \frac{(\partial_\theta a)(\partial_\theta b)}{a+b}\,.
\end{equation}
Thus, if the encoding is in only one of the parameters --- that is $\partial_\theta a = 0$ or $\partial_\theta b=0$ --- we get $\Fall{\tau} = 2 \Femp{\tau}$; 
The ED therefore contains only half of the full information available in the stochastic process. 
Finally, we mention that in this case of a 2-level system the ED has the same information as the sample mean; that is $\Femp{\tau} = \Favg{\tau}$. 

\subsection{Toy three-level system}\label{sec:examples_toy3}

To properly see a difference between $\Fall{\tau}$, $\Femp{\tau}$ and $\Favg{\tau}$ we must go to a system with at least three levels. 
The formulas however, become much more complicated in this case. 
We therefore study here a simple toy model where we parametrize the transition rates which allows us to gain some physical insights into what is happening. 

We consider a three-level system with states labeled by $-1$, $0$, and $+1$.
We assume transitions are only between neighboring states and described by the following transition matrix:  
\begin{equation}\label{eq:toy3_rateMat}
    \rateMat = \frac{\kappa}{1+\toyxi n} \left[ \begin{matrix}
    -\frac{1-\toyxi}{1-n} & \frac{1}{2} \frac{1-\toyxi}{1+n} & 0 \\
    \frac{1-\toyxi}{1-n} & -\frac{1+\toyxi n}{1-n^2} & \frac{1+\toyxi}{1+n} \\
    0 & \frac{1}{2} \frac{1+\toyxi}{1-n} & -\frac{1+\toyxi}{1+n}
    \end{matrix}\right]\,,
\end{equation}
with parameters $n\in(-1,1)$, $\toyxi\in(-1,1)$, and $\kappa>0$; $n$ and $\toyxi$ being unitless and $\kappa$ having units of frequency. 
A diagram of this model is shown in Fig.~\ref{fig:ex_setups}a.
The meaning of these parameters follows as below. 

First, $\kappa$ is an overall rate governing the time scale of the process; namely the dynamical activity which is the time-averaged rate at which transitions occur.
Next, the steady-state is 
\begin{equation}\label{eq:toy3_steady}
    \boldsymbol{p} = \left[ \begin{matrix}
        \left( \frac{1-n}{2} \right)^2 \\
        \frac{1}{2} \left( 1-n^2 \right) \\
        \left( \frac{1+n}{2} \right)^2
    \end{matrix} \right]\,,
\end{equation}
which depends only on $n$. 
In fact, the average of $x$ is 
\begin{equation}
    \mu = \sum_{x = -1,0,1} x p_x = n\,. 
\end{equation}
Finally, $\toyxi$ reflects the coupling between the states $\pm1$ and the state $0$. 
A simulation of the toy model is shown in Fig.~\ref{fig:toy3_summ}a using the Gillespie algorithm (see Appendix~\ref{app:gillespie}). 

\begin{figure*}[ht]
    \centering
    \includegraphics{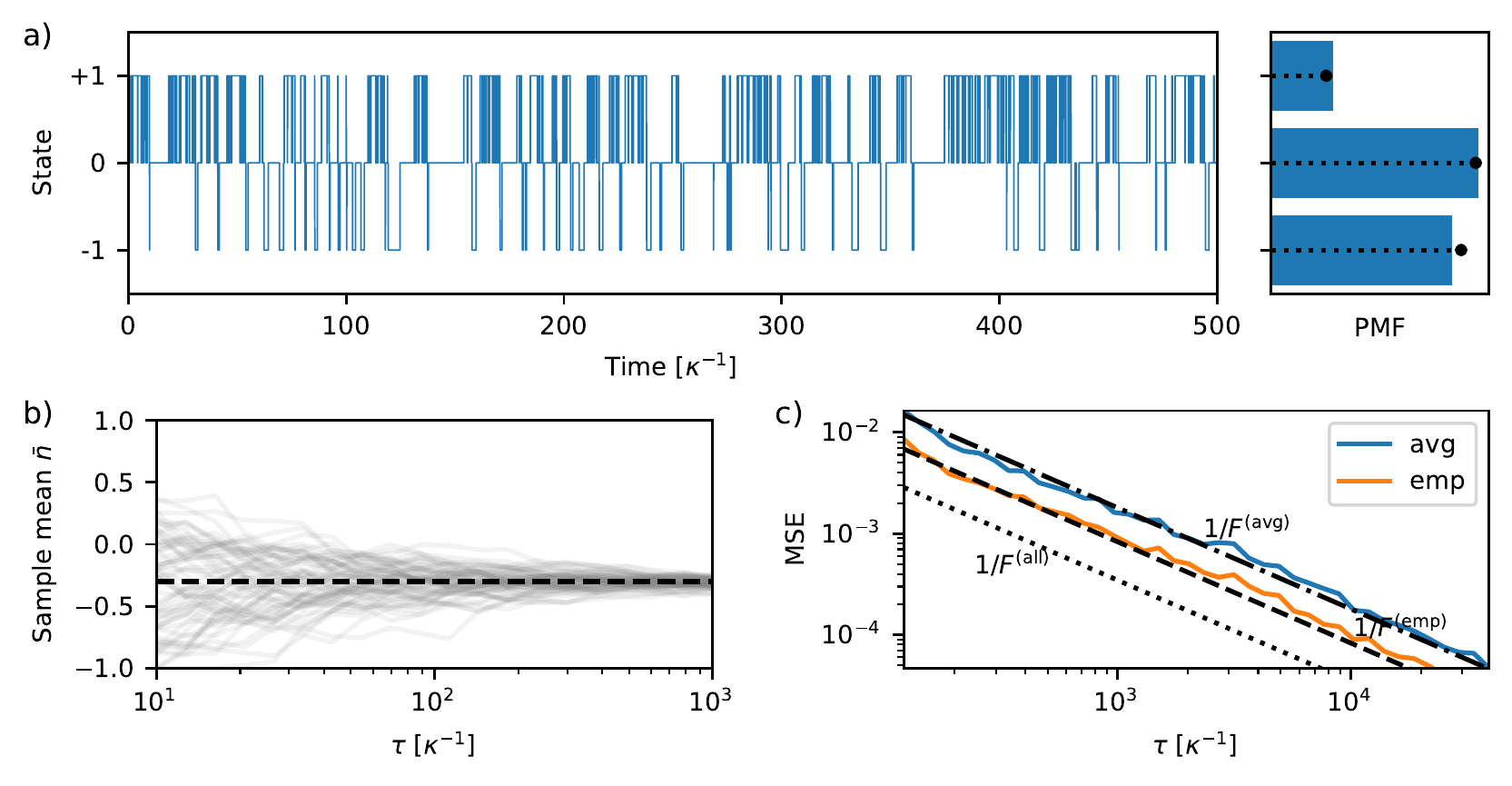}
    \caption{Numeric simulations of the three-level toy model with $n=-0.3$ and $\toyxi=0.75$. (a) A single stochastic trajectory. A histogram of the states is shown on the right along with the expected distribution (black points/lines) on the right. (b) The sample mean for many simulations and different simulation times $\tau$. A dashed line is drawn at $n$. (c) The mean-squared-error in the sample mean (blue) as well as with an estimator that uses the empirical distribution (orange): $\hat{n}_{\rm emp} = 2\sqrt{q_{+}}-1$. This is compared to the lower bound from Fisher information in the average state (Eq.~\eqref{eq:toy3_avgFish}, dot-dashed), empirical distribution (Eq.~\eqref{eq:toy3_empFish}, dashed), and in the full chain (Eq.~\eqref{eq:toy3_allFish}, dotted). Note that the estimator $\hat{n}_{\rm emp}$ using the empirical distribution may not be optimal, particularly with negative values of $\toyxi$.}
    \label{fig:toy3_summ}
\end{figure*}

Because the steady state [Eq.~\eqref{eq:toy3_steady}] is independent of $\toyxi$ and $\kappa$, only $n$ can be sensed with the empirical distribution. 
Using Eq.~\eqref{eq:Femp_master_DB}, the Fisher information for $\theta=n$ contained in the empirical distribution is
\begin{equation}\label{eq:toy3_empFish}
    \Femp{\tau}(n) = \frac{\kappa \tau}{\left( 1-n^2 \right)^2}\,.
\end{equation}
Conversely, the Fisher information in the full sequence is, from Eq.~\eqref{eq:Fall_Markov_tcont},
\begin{equation}\label{eq:toy3_allFish}
    \Fall{\tau}(n) = \left[ 1 + \left( \frac{n-\toyxi+2n^2\toyxi}{1+n\toyxi} \right)^2 \right] \Femp{\tau}(n)\,.
\end{equation}
This satisfies $\Fall{\tau}(n) \geqslant \Femp{\tau}(n)$.
Finally, the Fisher information from the average state follows from Eq.~\eqref{eq:avgFish_cMark}:
\begin{equation}\label{eq:toy3_avgFish}
    \Favg{\tau}(n) = \left( \frac{1-\toyxi^2}{1+n^2\toyxi^2} \right) \Femp{\tau}(n)\,,
\end{equation}
which is smaller than $\Femp{\tau}(n)$ and coincides with it only when $\toyxi = 0$.

\begin{figure}[ht]
    \centering
    \includegraphics{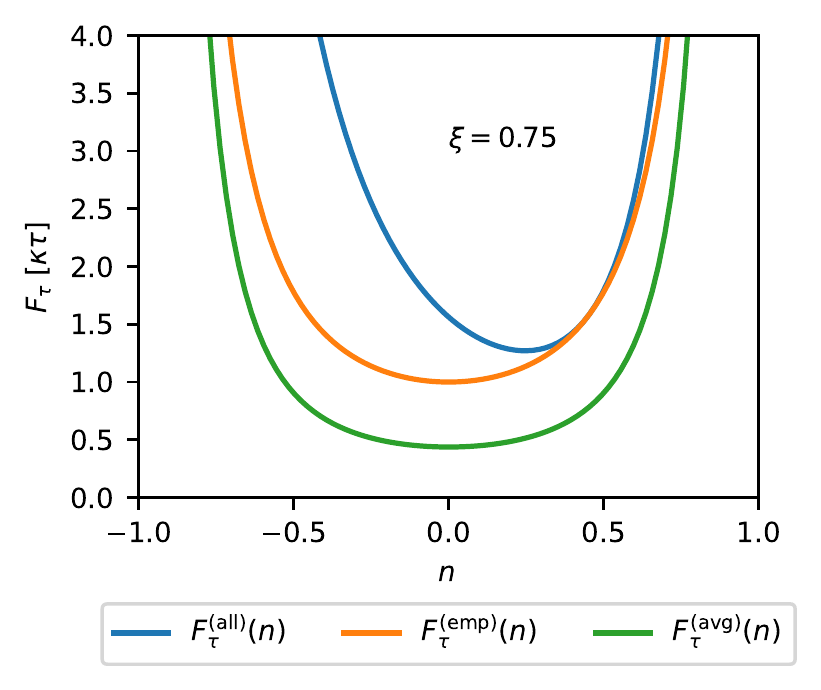}
    \caption{Comparing the Fisher information from different measurements derived from the three-level toy model described by Eq.~\eqref{eq:toy3_rateMat} with $\toyxi=0.75$. The Fisher information from the full chain (Eq.~\eqref{eq:toy3_allFish}, blue), the empirical distribution (Eq.~\eqref{eq:toy3_empFish}, orange), and the average state (Eq.~\eqref{eq:toy3_avgFish}, green). 
    }
    \label{fig:toy3_fisher}
\end{figure}

We illustrate the role played by these different quantities in in Fig.~\ref{fig:toy3_summ}.
In Fig.~\ref{fig:toy3_summ}b we illustrate an estimation protocol using the sample mean~\eqref{eq:sample_mean} as estimator. 
As can be seen, at the level of individual trajectories the simulations converge with increasing $\tau$ to the true value of $n$. 
In Fig.~\ref{fig:toy3_summ}c we plot the mean-squared error (MSE) from this estimation, averaged over many trajectories, as a function of $\tau$. 
We also include the corresponding bounds on the MSE from Eqs.~\eqref{eq:toy3_empFish}-\eqref{eq:toy3_avgFish}.
As can be seen, the sample mean (blue curve) is asymptotically bounded by $1/\Favg{\tau}(n)$. 
To go below this threshold one must employ estimators that use additional information. 
The orange curve in Fig.~\ref{fig:toy3_summ}c is one example, where the estimator $\hat{n}_{\rm emp} = 2 \sqrt{q_+} - 1$ includes information in the empirical distribution that is not contained in the sample mean.
The form of estimator is derived from the third entry of Eq.~\eqref{eq:toy3_steady}.
The MSE with $\hat{n}_{\rm emp}$ is shown to asymptotically converge to a value slightly above $1/\Femp{\tau}$. 
If we want to further go below the $1/\Femp{\tau}$ bound, then we would have to use an estimator that employs also the transition rates. 
The three quantities in Eqs.~\eqref{eq:toy3_empFish}-\eqref{eq:toy3_avgFish} are also compared in Fig.~\ref{fig:toy3_fisher}, as a function of $n$. 
As can be seen, the difference might be quite significant in some parameter regimes.

\subsection{Thermally-coupled harmonic oscillator}\label{sec:examples_QHO}

\begin{figure*}[ht]
    \centering
    \includegraphics{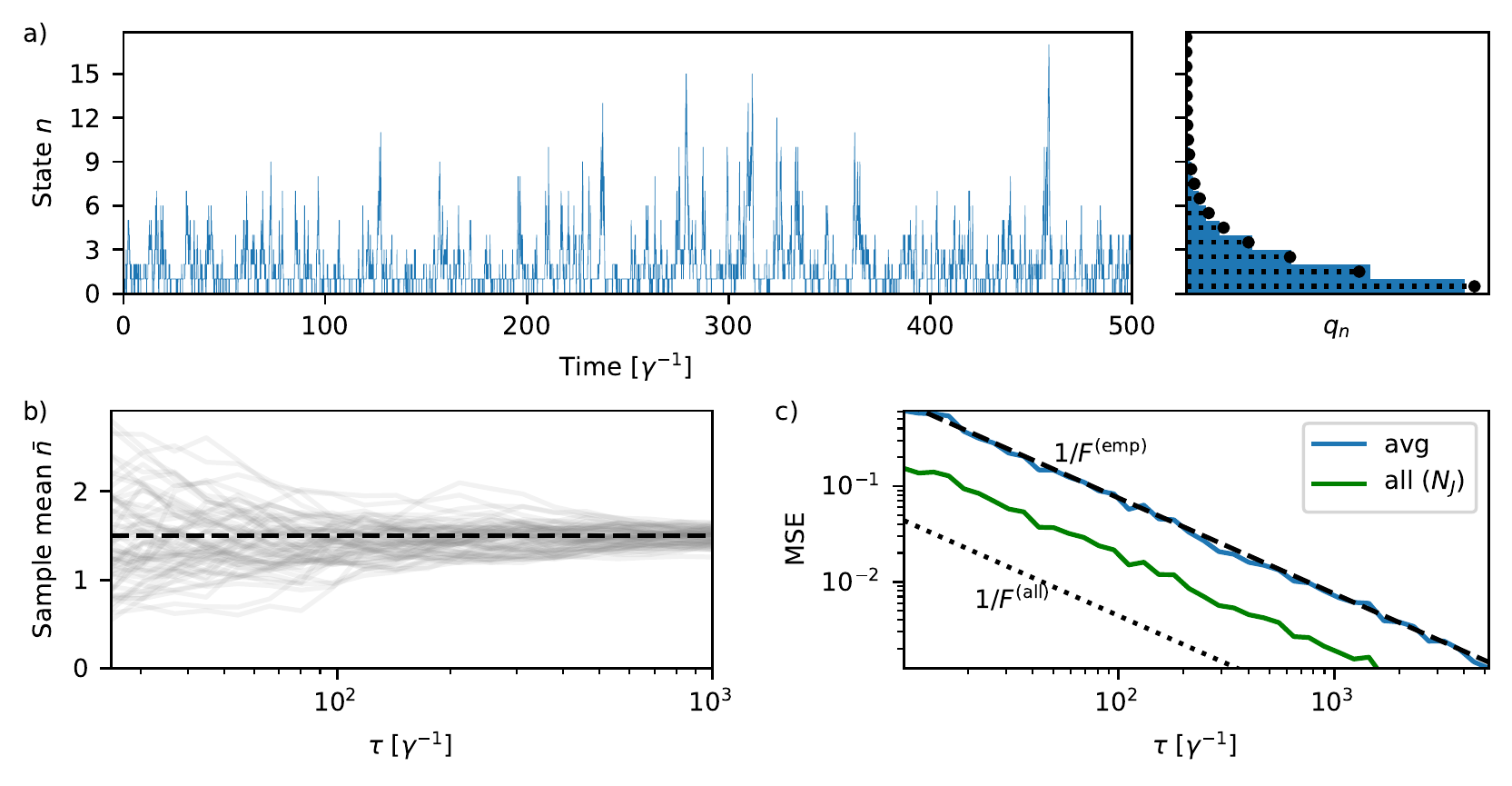}
    \caption{
    Stochastic dynamics of a quantum harmonic oscillator coupled to a thermal bath via Eqs.~\eqref{eq:master} and~\eqref{eq:ho_rateMat}, with $\Nb=1.5$. (a) A single simulated trajectory and the corresponding empirical distribution $q_n$. For sufficiently large times $q_n$ will converge to the steady-state distribution $p_n$ in Eq.~\eqref{eq:QHO_ss_distribution} (black points/lines). 
    (b) Sample mean as a function of $\tau$, which can be used as a simple estimator in this case. The dashed line denotes $\Nb=1.5$. 
    (c) In blue we show the mean-squared error of the sample mean as a function of $\tau$. 
    For long times this tends asymptotically to $1/\Femp{\tau}$ (dashed).
    The full precision $1/\Fall{\tau}$ (dotted) can only be reached with estimators making use of higher order statistics, beyond the empirical distribution. For example, in green, the number of jumps $N_J=2\gamma\tau \Nb\left(\Nb+1\right)$ --- information not contained in the empirical distribution --- provides an improved estimator.
    }
    \label{fig:tho_summ}
\end{figure*}

Next we consider a quantum harmonic oscillator coupled to a heat bath at a fixed temperature $T$. 
The system is described by  discrete energy levels given by the Fock states $\ket{n}$, for $n = 0,1,2,\ldots$, with energies $E_n = \hbar \omega (n+1/2)$ (Fig.~\ref{fig:ex_setups}b).
If the initial state is incoherent in the Fock basis, the dynamics in contact with the bath can be described by a classical Pauli master equation of the form~\eqref{eq:master}, where transitions only occur between neighboring energy states: 
\begin{equation}
    \frac{dp_n}{dt} = W_{n,n+1} p_{n+1} + W_{n,n-1} p_{n-1} - (W_{n+1,n} + W_{n-1,n}) p_n\,.
\end{equation}
The form of the transition rates can be determined from the corresponding Lindblad master equation for the quantum evolution~\cite{breuer2007}, and read
\begin{subequations}\label{eq:ho_rateMat}
    \begin{align}
        \rateMatTL_{n,n+1} &= \gamma \left( \Nb+1 \right)\left( n+1 \right) \\
        \rateMatTL_{n+1,n} &= \gamma \Nb\left( n+1 \right)\,,
    \end{align}
\end{subequations}
for $n=0,1,\ldots$. 
Here $\gamma >0$ is the coupling strength to the bath and $\Nb = (e^{\hbar \omega/T}-1)^{-1}$ is the Bose-Einstein distribution. 
The steady-state is the thermal distribution 
\begin{equation}\label{eq:QHO_ss_distribution}
    p_n = \left( 1-\frac{\Nb}{\Nb+1} \right)\left(\frac{\Nb}{\Nb+1}\right)^n = \left( 1-e^{-\hbar\omega\beta} \right) e^{-\hbar\omega\beta n}\,,
\end{equation}
and one can verify that it satisfies detailed balance [Eq.~\eqref{eq:detailed_balance}]. 
A stochastic trajectory of this process is shown in Fig.~\ref{fig:tho_summ}a.

\begin{figure}
    \centering
    \includegraphics{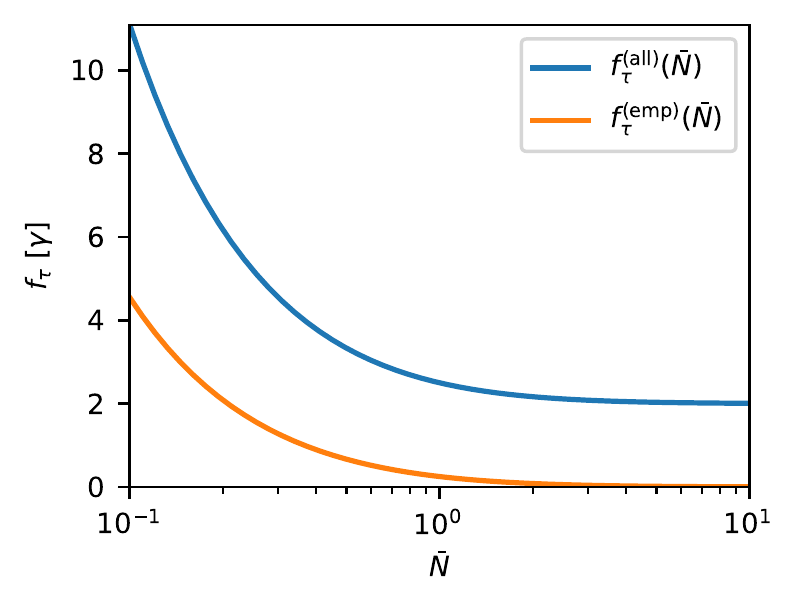}
    \caption{Fisher information rates $\fall = \Fall{\tau}/\tau$ and $\femp = \Femp{\tau}/\tau$ as a function of $\Nb$, for the quantum harmonic oscillator example, Eqs.~\eqref{eq:tho_empFish} and~\eqref{eq:tho_allFish}.
    }
    \label{fig:fisher_QHO}
\end{figure}

The steady-state [Eq.~\eqref{eq:QHO_ss_distribution}] is insensitive to $\gamma$. Hence, the only parameter we can do estimation on is $\Nb$. 
The Fisher information of $\Nb$ in the empirical distribution follows from Eq.~\eqref{eq:Femp_master_DB},
\begin{equation}\label{eq:tho_empFish}
    \Femp{\tau}(\Nb) = \frac{\gamma \tau}{2 \Nb \left( \Nb+1 \right)}\,.
\end{equation}
The Fisher information in the full sequence follows from Eq.~\eqref{eq:Fall_Markov_tcont}:
\begin{equation}\label{eq:tho_allFish}
    \Fall{\tau}(\Nb) = \gamma \tau \left( \frac{\Nb+1}{\Nb} + \frac{\Nb}{\Nb+1} \right) = 2\gamma \tau \cosh\left( \hbar\omega\beta \right)\,.
\end{equation}
We therefore see that the Fisher information rates (in units of $\gamma$) $\bar{f}^{\text{(all)}} = \Fall{\tau}/\gamma\tau$ and $\bar{f}^{\text{(emp)}} = \Femp{\tau}/\gamma\tau$ satisfy 
\begin{equation}
    \bar{f}^{\text{(all)}} = 2\bar{f}^{\text{(emp)}} + 2\,.
\end{equation}
The empirical distribution always loses at least half of the information. 
A plot of the two quantities is shown in Fig.~\ref{fig:fisher_QHO}. 

In Fig.~\ref{fig:tho_summ}(b) we show an example of how to estimate $\Nb$ from the outcomes. An obvious choice of estimator in this case is the sample mean [Eq.~\eqref{eq:sample_mean}] which, from Eq.~\eqref{eq:QHO_ss_distribution}, should yield precisely $\Nb$. 
The mean-squared error between the sample mean and the true $\Nb$ is shown in Fig.~\ref{fig:tho_summ}(c). 
As can be seen, the error saturates to $1/\Femp{\tau}$.
It turns out that in this case $\Favg{\tau}$ (computed numerically using Eq.~\eqref{eq:avgFish_cMark}) is the same as $\Femp{\tau}$. We attribute this to a special coincidence of this model, as it is something which is clearly violated in others, such as the three-level maser discussed in Sec.~\ref{sec:examples_maser}. 
To build an estimator that can improve upon the $1/\Femp{\tau}$ MSE bound we must use information related to the transition rates. 
One example of such an estimator is the dynamical activity
\begin{equation}
    \bar{R} = \sum_{j\neq n} W_{nj} p_j = 2\gamma \Nb \left(\Nb+1\right)\,,
\end{equation}
which uses both the empirical distribution and the transition rates. An example of the mean-squared error in this case is shown in Fig.~\ref{fig:tho_summ}(c) in green. 
The error lies between the $1/\Femp{\tau}$ and $1/\Fall{\tau}$ bounds. A more complicated estimator that incorporates additional information, e.g., between which states transitions are occurring, is expected to push the error closer to the $1/\Fall{\tau}$ bound.

\subsection{Three-level maser}\label{sec:examples_maser}
A maser can be modeled as a three-level system with states $\{ \ket{1}, \ket{2}, \ket{3} \}$ and corresponding distinct energies (in ascending order): $E_1$, $E_2$, and $E_3$. 
The allowed transitions are shown in Fig.~\ref{fig:ex_setups}c.
We consider here the incoherent implementation of the three-level maser:
a  hot thermal bath, with coupling $\gamma_h$ and Bose-Einstein distribution $\sim \Nb_h$, acting  between states $\ket{1}$ and $\ket{3}$; 
a cold thermal bath with coupling $\gamma_c$ and Bose-Einstein distribution $\sim \Nb_c$, acting between $\ket{3}$ and $\ket{2}$;
and an incoherent spontaneous emission of a photon into an optical mode with rate $\gamma_w$. 
The rate matrix reads
\begin{equation}
    \rateMatTL = \left[ \begin{matrix}
          & \gamma_w & \gamma_h \left( \Nb_h+1 \right) \\
        0 &   & \gamma_c \left( \Nb_c+1 \right) \\
        \gamma_h \Nb_h & \gamma_c \Nb_c &  
    \end{matrix} \right]\,.
\end{equation}
The steady state is then
\begin{align}\label{eq:maser_rateMat}
    \boldsymbol{p} &= 
    \frac{1}{\gamma_c \gamma_h \left(3 \Nb_c \Nb_h+\Nb_c+\Nb_h\right)+\gamma_w \left(\gamma_c \left(\Nb_c+1\right)+\gamma_h \left(2 \Nb_h+1\right)\right)} \nonumber \\
    &\quad\times
    \left[\begin{matrix}
        {\gamma_c \gamma_h \Nb_c \left(\Nb_h+1\right)+\gamma_w \left[\gamma_c \left(\Nb_c+1\right)+\gamma_h \left(\Nb_h+1\right)\right]} \\
        {\gamma_c \gamma_h \left(\Nb_c+1\right) \Nb_h} \\
        {\gamma_c \gamma_h \Nb_c \Nb_h+\gamma_w \gamma_h \Nb_h}
    \end{matrix}
    \right]\,.
\end{align}
Detailed balance [Eq.~\eqref{eq:detailed_balance}] is violated whenever $\gamma_w\neq0$.

To simplify the number of free parameters we define $\frac{\gamma_h}{\gamma_w} = \frac{\gamma_c}{\gamma_w} = \gamma$.
We also set $\Nb_c \to 0$. 
A key feature of the three-level maser is population inversion between $\ket{2}$ and $\ket{1}$. It occurs when
\begin{equation}
    \Nb_h > \frac{2}{\gamma-1} + \frac{\gamma+1}{\gamma-1} \Nb_c\,,
\end{equation}
provided $\gamma>1$. No population inversion occurs for $\gamma<1$.

\begin{figure}[ht]
    \centering
    \includegraphics{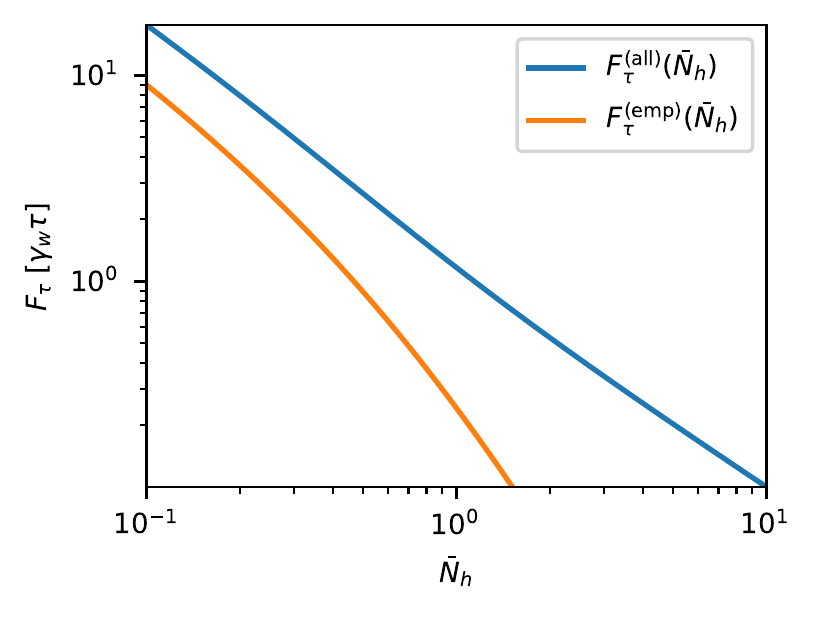}
    \caption{Comparing the Fisher information from different measurements derived from the three-level maser described by Eq.~\eqref{eq:maser_rateMat} with $\gamma_h=\gamma_c=2\gamma_w$ and $\Nb_c\to0$. The Fisher information is shown from the full chain (blue, Eq.~\eqref{eq:maser_allFish}) and the empirical distribution (orange, Eq.~\eqref{eq:maser_empFish}). }
    \label{fig:maser_fisher}
\end{figure}

As with the previous examples, the Fisher information can be calculated with respect to the parameters of the theory as before using Eq.~\eqref{eq:Fall_Markov_tcont}, Eq.~\eqref{eq:Femp_master_no_DB}, and Eq.~\eqref{eq:avgFish_cMark}. Because detailed balance does not hold, one cannot use Eq.~\eqref{eq:Femp_master_DB} for the ED. 
For concreteness, we focus here on the Fisher information of $\Nb_h$; that is, we frame this as a thermometry problem for the temperature of the hot bath.
The results are:
\begin{align}
    \Fall{\tau}(\Nb_h)&= \frac{\gamma \tau}{\Delta} \frac{2 \Nb_h^2+3 \Nb_h+2}{\Nb_h+1}, \label{eq:maser_allFish}\\
    \Femp{\tau}(\Nb_h)&= \frac{\gamma \tau}{\Delta} \frac{8}{4 \Nb_h+7}\,,\label{eq:maser_empFish}
\end{align} 
where $\Delta = \Nb_h \left((\gamma +2) \Nb_h+2\right)$.
We compare them Fig.~\ref{fig:maser_fisher}.
For large $\Nb_h$, we have the asymptotic scalings
\begin{equation}
\Fall{\tau} \propto \frac{1}{\Nb_h}\,, 
\qquad 
\Femp{\tau} \propto \frac{1}{\Nb_h^3}\,,
\end{equation}
so that the discrepancy with $\Fall{\tau}$ becomes significant. 
Conversely, for small $\Nb_h$ we get 
\begin{equation}
    \Fall{\tau} \simeq 2 \gamma\tau\,, 
    \qquad 
    \Femp{\tau} \simeq \frac{8\gamma\tau}{7}\,.
\end{equation}

\subsection{Spin-one chain thermometry}\label{sec:examples_spin_chain}

We now turn to an example that is not a continuous-time master equation. 
A spin-one classical chain can be modeled as a sequence of $N$ discrete random variables $s_j$ (spins), each taking values in $\{-1, 0, 1\}$. The spins interact  with each other and with an external magnetic field $B$ according to the classical Hamiltonian
\begin{equation}
    H(s_{1:N}) = - B\sum_{j} s_j - J\sum_{j} s_j s_{j+1}\,,
\end{equation}
where $J$ is the coupling parameter, and the summation implicitly yields periodic boundary conditions. In particular, a thermal state of the chain has probability distribution
\begin{equation}
   \probof{s_{1:N}} = \frac{e^{-H(s_{1:N})/T}}{Z}\,,
\end{equation}
where the Boltzmann constant has been set to 1, $T$ is the temperature and $Z$ the partition function. In the thermodynamic limit $N\to\infty$, the model can be studied by exploiting the transfer matrix formalism~\cite{Kramers_1941, Feldman_1998b, Feldman_1998}. 
For details see Ref.~\cite{radaelli_fisher_2022}.

Let us consider a sequential access to the spin chain; the spins are measured one by one, in sequence, from left to right. The obtained information can be either kept in full, or compressed in terms of the empirical distribution (the count of $0, \pm 1$ spins) or even further via the sample mean. The compression is, as expected, associated to a reduction of the Fisher information, and consequently to an increase of the estimation error.

\begin{figure}[htbp]
    \centering
    \includegraphics[width=\columnwidth]{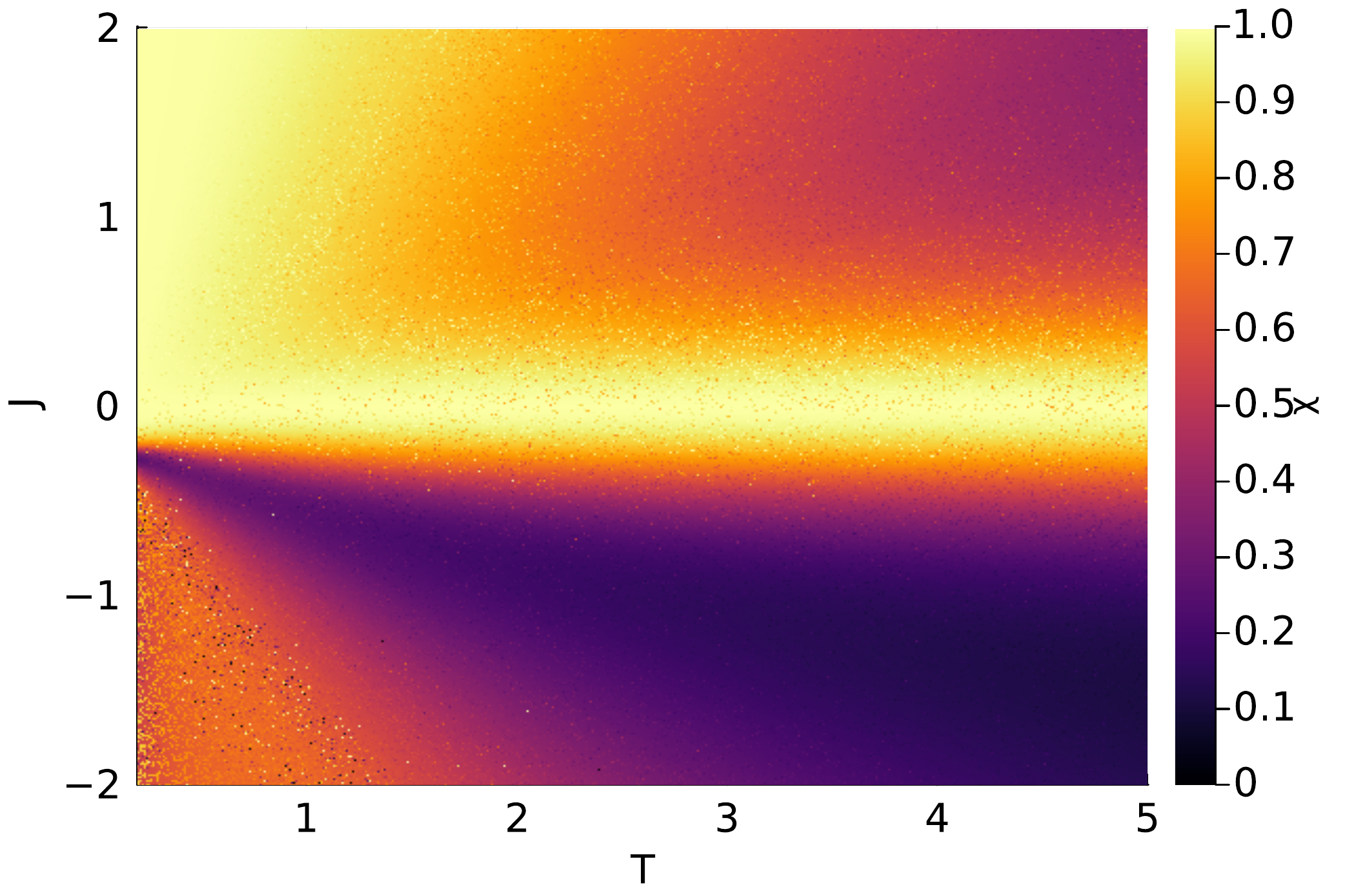}
    \caption{Heatmap of $\chi = \Femp{N}/\Fall{N}$ on a spin-one chain, for different values of $T$ and $J$. Note the difference in the behavior between ferromagnetic and anti-ferromagnetic regimes, for external magnetic field $B=0.5$.}
    \label{fig:heatmap_chi}
\end{figure}

While the inequality \eqref{eq:3_fisher_inequalities} fixes a clear hierarchy of decreasing precision for increasing compression, different regimes of the spin chain system can yield vastly different behaviours. In Figure~\ref{fig:heatmap_chi}, we represent the ratio $\chi$ between $\Femp{N}$ and $\Fall{N}$ for different values of $T$ and $J$ (given \eqref{eq:3_fisher_inequalities} and the positivity of the Fisher information, $0 \leq \chi \leq 1$). 
By definition, the empirical distribution is only able to collect information contained in the stationary distribution, while the full statistic also harvests information from the correlations. In the ferromagnetic regime $J \geq 0$, where the information about the temperature is mostly encoded in the stationary distribution~\cite{radaelli_fisher_2022}, the loss of precision due to the empirical distribution compression is negligible ($\chi \sim 1$). On the contrary, in the anti-ferromagnetic case the information is mostly contained in the correlations, hence the empirical distribution turns out to be highly ineffective ($\chi \sim 0$).

\section{Conclusions}\label{sec:conclusions}

In this paper we study parameter estimation with stochastic processes. 
Unlike standard settings, we focus specifically on data that are obtained from processes which are not iid, but depend on previous outcomes. 
These temporal correlations can affect the rate at which we acquire information about the parameter being estimated. 
We derive several expressions quantifying the information that remains in different relevant functions (estimators) computed from the full data. 
In particular, we consider the Fisher information contained in the empirical distribution. Intuitively, this is the distribution one obtains by making a histogram for the data obtained from a stochastic process. 
Our main results are summarized in Table~\ref{tab:results_summary}. 
Each column represents a level of ``compression,'' from the full process to the empirical distribution to the sample mean, and each row represents a class of stochastic processes: from a general stationary process to discrete-time Markov chains to continuous-time master equations.

In realistic settings, it may be impractical to build estimators for a desired parameter that make use of all possible aspects of the data. 
Instead, what is usually done in practice is to compress the outcome, e.g., either as the empirical distribution (a histogram) or the sample mean. 
Our results provide the fundamental loss of information that is incurred in this process by describing a new lower bound on the parameter's uncertainty when less information is available. 
We therefore believe our results will be useful for researchers exploring the metrological power of stationary stochastic processes.

\section*{Acknowledgments}

The authors acknowledge fruitful discussions with M.~Mitchison, M.~Kewming, S.~Campbell and A.~Kiely. MR acknowledges funding by the Irish Research Council under Government of Ireland Postgraduate Scheme grant number GOIPG/2022/2321. FCB acknowledges funding by the Irish Research Council under grant number IRCLA/2022/3922.

\bibliography{references}

\appendix

\begin{widetext}

\section{Covariance matrix of the empirical distribution}
\label{app:covariance_ED}

In this appendix we prove Eq.~\eqref{eq:ED_covariance_matrix} for the covariance matrix of the empirical distribution~\eqref{eq:empDist}.
From the definition of the covariance matrix we have 
\begin{equation}
\begin{aligned}
    \Sigma_{xy} 
    &= {\rm Cov}(q_x,q_y) 
    \\
    &= \frac{1}{N^2} \sum_{i,j} {\rm Cov}\left(\delta_{X_i,x}, \delta_{X_j,y}\right)
    \\
    &= \frac{1}{N^2} \sum_i {\rm Cov}\left(\delta_{X_i,x}, \delta_{X_i,y}\right)
    + \frac{1}{N^2} \sum_i\sum_{j>i}{\rm Cov}\left(\delta_{X_i,x}, \delta_{X_j,y}\right)
     + \frac{1}{N^2} \sum_i \sum_{j<i} {\rm Cov}\left(\delta_{X_i,x}, \delta_{X_j,y}\right)\,.
\end{aligned}
\end{equation}
We now use the fact that the stochastic process is stationary. In the first sum, all $N$ terms will be equal. 
In the second and third terms, ${\rm Cov}\left(\delta_{X_i,x}, \delta_{X_j,y}\right)$ depends only on the distance $i-j$. 
We may then use an identity from combinatorics
\begin{equation}
    \sum_{i=1}^{N-1} \sum_{j=i+1}^N C_{j-i} = \sum_{k=1}^{N-1} (N-k) C_k\,,
\end{equation}
valid for any function $C_k$. 
This then yields
\begin{equation}\label{eq:app_covar_unsimp2}
    \Sigma_{xy} = \frac{1}{N} {\rm Cov}\left(\delta_{X_1,x}, \delta_{X_1,y}\right)
    + 
    \frac{1}{N} \sum_{k=1}^{N-1} (1-k/N) {\rm Cov}\left(\delta_{X_1,x}, \delta_{X_{k+1},y}\right)
    + 
    \frac{1}{N} \sum_{k=1}^{N-1} (1-k/N){\rm Cov}\left(\delta_{X_{k+1},x}, \delta_{X_1,y}\right)\,.
\end{equation}

We can next compute all terms explicitly. 
First, 
\begin{equation}
    \begin{aligned}
    {\rm Cov}\left(\delta_{X_1,x}, \delta_{X_1,y}\right)
    &= \expect{\delta_{X_1,x}\delta_{X_1,y}} - \expect{\delta_{X_1,x}} \expect{\delta_{X_1,y}}
    \\
    &= p_x \delta_{x,y} - p_x p_y\,.
    \end{aligned}
\end{equation}
And second: 
\begin{equation}
\begin{aligned}
    {\rm Cov}\left(\delta_{X_1,x}, \delta_{X_{k+1},y}\right) &=
    \probof{X_1=x,X_{k+1}=y} - p_x p_y
    \\ &=\left[\probgiven{X_{k+1}=y}{X_1=x} -p_y\right]p_x
    \\
    &= \left[p_{y\gets x}(k) - p_y\right] p_x\,,
\end{aligned}
\end{equation}
where 
\begin{equation}
p_{x\gets y}(k) := \probgiven{X_{i+k}=x}{X_{i}=y}\,.
\end{equation}
With this expression in mind, one can define a matrix that encodes the correlations between measurements,
\begin{equation}\label{eq:app_Psi_matrix}
    \Psi_{xy} = \sum_{k=1}^{N-1} \left(1-\frac{k}{N}\right) \Big[p_{x\gets y}(k) - p_x\Big]\,.
\end{equation}
Plugging these results in Eq.~\eqref{eq:app_covar_unsimp2} then yields the final formula for $\Sigma$ in Eq.~\eqref{eq:ED_covariance_matrix} of the main text:
\begin{equation}\label{eq:app_ED_covariance_matrix}
    \Sigma_{xy} = \frac{1}{N} \left( p_x (\delta_{xy}- p_y) + \Psi_{xy} p_y + \Psi_{yx} p_x
    \right)\,.
\end{equation}

\section{Fisher information of the empirical distribution}
\label{app:F_emp_computation}

In this appendix we prove the first main result of our paper, Eq.~\eqref{eq:Femp_general_stoch_process}, for the Fisher information of the empirical distribution $\boldsymbol{q}$. 
The result holds for large $N$ and follows from the fact that, in this limit, the components of the empirical distribution $q_x$ are essentially random variables sampled from a joint multivariate Gaussian distribution with mean $p_x$ (the steady-state distribution) and covariance matrix $\Sigma_{xy}$ [Eq.~\eqref{eq:ED_covariance_matrix} and Eq.~\eqref{eq:app_ED_covariance_matrix} in Appendix~\ref{app:covariance_ED}] under the appropriate conditions and approximations. 
The Fisher information of a multi-variate Gaussian with mean vector $\bm{\mu}$ and covariance matrix $\Sigma$ is 
\begin{equation}\label{F_Gaussian}
    F^{\rm (Gauss)} = (\partial_\theta \bm{\mu})\trans \Sigma^{-1} (\partial_\theta \bm{\mu}) + \frac{1}{2} \tr\Big\{ \Sigma^{-1} (\partial_\theta \Sigma) \Sigma^{-1} (\partial_\theta \Sigma)\Big\}\,.
\end{equation}
In our case the covariance matrix~\eqref{eq:ED_covariance_matrix} scales as $\Sigma = \tilde{\Sigma}/N$. Hence, the first term will be $\mathcal{O}(N)$ while the second will be $\mathcal{O}(1)$. 
The leading term in $\Femp{N}$ will therefore be 
\begin{equation}\label{F_Gaussian_2}
    \Femp{N} = N (\partial_\theta \bm{p})\trans \tilde{\Sigma}^{-1} (\partial_\theta \bm{p})\,.
\end{equation}

The covariance matrix~\eqref{eq:ED_covariance_matrix} can be written in matrix form as 
\begin{equation}\label{app_Sigma_matrix_form}
    \tilde{\Sigma} = \mathbb{P} + U - \bm{p}\bm{p}\trans\,, 
    \qquad 
    U := \Psi \mathbb{P} + \mathbb{P}\Psi\trans\,,
\end{equation}
where $\mathbb{P} = {\rm diag}(p_x)$ and
$\Psi$, defined in Eq.~\eqref{eq:app_Psi_matrix}, encodes correlations between measurements. The last term is the outer product with elements $(\bm{p} \bm{p}\trans)_{xy} = p_x p_y$. 
When trying to insert Eq.~\eqref{app_Sigma_matrix_form} into Eq.~\eqref{F_Gaussian_2}, however, a complication arises. 
Namely  because $\sum_x q_x = 1$, the $d$ random variables $q_x$ are not all statistically independent, causing the matrix $\Sigma$ (or $\tilde{\Sigma}$) to be singular.
This can be avoided by working with $d-1$ elements of $q_x$. 
However, the manipulations become somewhat cumbersome. 
A simpler approach is to add an infinitesimal amount of noise in the $q_x$ by defining a new empirical distribution $q_x + \epsilon_x/\sqrt{N}$, where $\epsilon_x$ are arbitrary independent random variables satisfying $E[\epsilon_x] = 0$ and ${\rm Var}(\epsilon_x) \neq 0$. 
This changes the covariance matrix to $\tilde{\Sigma} + \mathcal{E}$, where $\mathcal{E}$ is a diagonal matrix with  $\mathcal{E}_{xx} = {\rm Var}(\epsilon_x)$. 
We may then use $\left(\tilde{\Sigma} + \mathcal{E}\right)^{-1}$ in Eq.~\eqref{F_Gaussian_2} and, in the end, take the limit $\mathcal{E}\to 0$ (by which we mean the limit where all entries of $\mathcal{E}$ tend to zero). 

As we show below in Appendix~\ref{app:proof_inverse}, for small $\mathcal{E}$ the following expansion holds:
\begin{align}\label{Sigma_epsilon_matrix_relation}
    (\tilde{\Sigma} + \mathcal{E})^{-1} &= (\mathbb{P} + U)^{-1} 
    - \frac{(\mathbb{P}+U)^{-1} \mathcal{E}\bm{u}\bm{u}\trans + \bm{u} \bm{u}\trans \mathcal{E} (\mathbb{P} + U)^{-1}}{\bm{u}\trans \mathcal{E} \bm{u}} + \frac{\bm{u}\bm{u}\trans}{\bm{u}\trans \mathcal{E}\bm{u}} + \mathcal{O}(\mathcal{E})\,,
\end{align}
where $\bm{u} = [1,\ldots,1]\trans$ is a column vector with all entries equal to 1. 
The third term in Eq.~\eqref{Sigma_epsilon_matrix_relation} diverges in the limit $\mathcal{E} \to 0$, while the first two terms are both of order $\mathcal{O}(1)$ in $\mathcal{E}$. 
However, the diverging terms vanish when ``sandwiched'' between vectors whose components add up to zero; i.e., that are perpendicular to $\boldsymbol{u}$. The normalization of the probabilities $\sum_x p_x=1$ implies $\sum_x \partial_\theta p_x=0$. Therefore, inserting~\eqref{Sigma_epsilon_matrix_relation} in Eq.~\eqref{F_Gaussian_2} causes both the second and third terms to vanish, leading to 
\begin{equation}
    \Femp{N} = N (\partial_\theta\bm{p})\trans (\mathbb{P} + U)^{-1} (\partial_\theta \bm{p}) + \mathcal{O}(\mathcal{E})\,. 
\end{equation}
In the limit $\mathcal{E}\to0$ we therefore recover our first main result, Eq.~\eqref{eq:Femp_general_stoch_process}.
Notice how this result is independent of the nature or structure of the noise matrix $\mathcal{E}$. 
Any infinitesimal noise will therefore work.

\section{Proof of Eq.~\eqref{Sigma_epsilon_matrix_relation}}
\label{app:proof_inverse}

The proof relies on two matrix identities:
the series expansion 
\begin{equation}\label{eq:app_matrix_inverse_series}
    (A+B)^{-1} = A^{-1} - A^{-1} B A^{-1} + \cdots + \left(-1\right)^k A^{-1}\left(BA^{-1}\right)^k + \cdots
\end{equation}
and the Sherman-Morrison formula, 
\begin{equation}\label{eq:sherman_morrison}
    (X - \bm{w} \bm{v}\trans)^{-1} = X^{-1} + \frac{X^{-1} \bm{w} \bm{v}\trans X^{-1}}{1 - \bm{v}\trans X^{-1} \bm{w}}\,. 
\end{equation}
We first use these two relations to  establish the following identity. 
Let $X = A + \mathcal{E}$, with $A$ invertible. We will also be interested in the case that  $\bm{v}\trans A^{-1} \bm{w} = 1$. 
First, applying Eq.~\eqref{eq:sherman_morrison} yields
\begin{equation}
    (A + \mathcal{E} - \bm{w} \bm{v}\trans)^{-1} = \left(A + \mathcal{E}\right)^{-1} + \frac{\left(A + \mathcal{E}\right)^{-1} \bm{w} \bm{v}\trans \left(A + \mathcal{E}\right)^{-1}}{1 - \bm{v}\trans \left(A + \mathcal{E}\right)^{-1} \bm{w}}\,.
\end{equation}
The next step would be to use Eq.~\eqref{eq:app_matrix_inverse_series} to expand the above equation and omit $\mathcal{O}(\mathcal{E})$ terms that will vanish when taking $\mathcal{E}\to 0$. One must be careful when omitting terms from the fraction because the assumption that $\bm{v}\trans A^{-1} \bm{w} = 1$ implies that the denominator is $\mathcal{O}(\mathcal{E})$, so one must retain the same order terms in the numerator. 
Then a series expansion in powers of $\mathcal{E}$ yields, to leading order, 
\begin{equation}\label{eq:app_new_relation_A}
    (A + \mathcal{E} - \bm{w}\bm{v}\trans)^{-1} = A^{-1} - \frac{
    A^{-1} \mathcal{E} A^{-1} \bm{w} \bm{v}\trans A^{-1} + A^{-1} \bm{w} \bm{v}\trans A^{-1} \mathcal{E} A^{-1}
    }{\bm{v}\trans A^{-1} \mathcal{E} A^{-1} \bm{w}} + \frac{A^{-1} \bm{w}\bm{v}\trans A^{-1}}{\bm{v}\trans A^{-1} \mathcal{E}A^{-1} \bm{w}} + \mathcal{O}(\mathcal{E})\,.
\end{equation}

We now apply our case to Eq.~\eqref{eq:app_new_relation_A}, where $A = \mathbb{P} + U$ and  $\bm{w} = \bm{v} = \bm{p}$ is the steady-state distribution.
One can verify from the definition of $\Psi$ in Eq.~\eqref{eq:app_Psi_matrix} that $\bm{u}\trans \Psi = \Psi\trans \bm{u} = 0$ and $\Psi \bm{p} = \bm{p} \Psi\trans = 0$. 
This, together with the fact that $\mathbb{P} \bm{u} = \bm{p}$, implies that $U \bm{u} = \bm{u}\trans U = 0$. 
Finally, since $\mathbb{P}^{-1} \bm{p} = \bm{u}$, Eq.~\eqref{eq:app_matrix_inverse_series} yields 
\begin{equation}
    A^{-1}\bm{w} \equiv (\mathbb{P} + U)^{-1} \bm{p} = \left(\mathbb{P}^{-1}-\mathbb{P}^{-1} U \mathbb{P}^{-1} +\ldots\right) \bm{p} = \bm{u} - \mathbb{P} U \bm{u} +\mathbb{P}^{-1} U \mathbb{P}^{-1} U \bm{u} + \ldots = \bm{u}\,.
\end{equation}
It immediately follows that $\bm{v}\trans A^{-1}\bm{w} = 1$, satisfying the assumptions. Likewise, $A$ is symmetric, so $\bm{p}\trans A^{-1}=\bm{u}\trans$, so one finds
\begin{equation}
    (\mathbb{P} + U - \bm{p}\bm{p}\trans + \mathcal{E})^{-1} = \left(\mathbb{P} + U\right)^{-1} 
    - \frac{\left(\mathbb{P} + U\right)^{-1} \mathcal{E} \bm{u} \bm{u}\trans + \bm{u} \bm{u}\trans \mathcal{E} \left(\mathbb{P} + U\right)^{-1}}{\bm{u}\trans \mathcal{E} \bm{u}} 
    + \frac{ \bm{u}\bm{u}\trans}{\bm{u}\trans  \mathcal{E}\bm{u}} + \mathcal{O}(\mathcal{E})\,.
\end{equation}
Recalling $\tilde{\Sigma}=\mathbb{P} + U - \bm{p}\bm{p}\trans$ yields Eq.~\eqref{Sigma_epsilon_matrix_relation}.

\end{widetext}

\section{Properties of the Drazin inverse} \label{app:drazin}

Here, we consider some useful properties of a generalization of the matrix inverse used for diagonalizable matrices: the Drazin inverse. The idea behind this inverse is that a diagonalizable matrix can be written in the form
\begin{equation}\label{eq:projective_decomp}
    A = \sum_{i} \lambda_i P_i\,,
\end{equation}
where $\{\lambda_i\}$ are the eigenvalues associated with the right-eigenvectors $\boldsymbol{x_i}$ and left-eigenvectors $\boldsymbol{y_i}\trans$, and $\{P_i\}$ are projective matrices with the properties:
\begin{subequations}\label{eq:projMat}
    \begin{align}
        P_i &= \boldsymbol{x_i}\boldsymbol{y_i}\trans\,, \\
        P_iP_j &= P_i \delta_{ij}\,,\\
        \sum_i P_i &= \ident\,. \label{eq:projMat_normal}
    \end{align}
\end{subequations}
When writing the matrix in the form of Eq.~\eqref{eq:projective_decomp}, the inverse is obtained by simply replacing $\lambda_i$ with $\lambda_i^{-1}$. However, this fails when $\lambda_j=0$ for some values of $j$; this is precisely the case for singular matrices. Of course, terms with $\lambda_j=0$ do not appear in the sum in Eq.~\ref{eq:projective_decomp}, so one may be tempted to simply ignore these terms, resulting in the Drazin inverse:
\begin{equation}
    A^\draz = \sum_{i,\lambda_i\neq 0} \frac{1}{\lambda_i} P_i\,.
\end{equation}
This is not a proper matrix inverse because it does not act properly on the nullspace of $A$; i.e., $A^\draz A\boldsymbol{v}\neq \boldsymbol{v}$ when $\boldsymbol{v}$ has some component in the kernel of $A$. 

Some properties of the Drazin inverse immediately follow from the definition:
\begin{align}\label{eq:draz_conj}
    \left( A^\draz \right)^\draz &= A\,, \\
    \left( A^\draz \right)\trans &= \left( A\trans \right)^\draz\,, \\
    \text{for }A\text{ invertible, } A^{-1} &= A^\draz\,.
\end{align}
Additionally, the Drazin inverse of $A$ conjugated with an invertible matrix $B$ is
\begin{equation}
    BA^\draz B^{-1} = \left( BAB^{-1} \right)^\draz\,,
\end{equation}
which will be proven as follows. Writing out the conjugation with Eq.~\eqref{eq:projective_decomp}, one finds
\[
    BAB^{-1} = \sum_i \lambda_i BP_iB^{-1}\,.
\]
It is straightforward to show that $BP_iB^{-1}$ preserves all the properties of projective matrices described by Eq.~\eqref{eq:projMat}. It follows that the Drazin inverse of $BAB^{-1}$ is
\begin{align*}
    \left( BAB^{-1} \right)^\draz &= \sum_{i,\lambda_i\neq0} \frac{1}{\lambda_i} BP_iB^{-1} \\
    &= B \left( \sum_{i,\lambda_i\neq0} \frac{1}{\lambda_i} P_i \right) B^{-1} \\
    &= BA^\draz B^{-1}\,.
\end{align*}

Before deriving a few more properties of the Drazin inverse, it will help to define a projection operator into the (co)kernel of a matrix. Using the same decomposition from Eq.~\eqref{eq:projective_decomp}, one can define the null projection operator for some matrix $A$ by
\begin{equation}
    N_A = \sum_{j,\lambda_j=0} P_j\,.
\end{equation}
A few immediate properties arise:
\begin{subequations}
    \begin{align}
        N_AN_A &= N_A\,, \label{eq:nullProj_idemp}\\
        N_AA&=AN_A\,, \label{eq:nullProj_annih}\\
        N_A^\draz=N_{A^\draz}&=N_A\,, \\
        \left(A+N_A\right)^{-1} &= A^\draz + N_A\,, \\
        A^\draz A = AA^\draz &= \ident - N_A\,.
    \end{align}
\end{subequations}
If follows from Eq.~\eqref{eq:nullProj_idemp} that $N_A$ is a projection operator and from Eq.~\eqref{eq:nullProj_annih} that it annihilates vectors and covectors in the image and coimage of $A$ --- i.e., it projects vectors and covectors into the kernel and cokernel\footnote{The cokernel is identified as the set of covectors that annihilate the image of $A$. Likewise, the coimage of $A$ is the set of covectors that can be written in the form $\boldsymbol{y}\trans A$.} of $A$, respectively. To see this, let $\boldsymbol{v}$ be a vector in the image of $A$ and $\boldsymbol{u}\trans$ a vector in the coimage of $A$. Thus, there exists a vector $\boldsymbol{x}$ such that $\boldsymbol{v}=A\boldsymbol{x}$ and a covector $\boldsymbol{y}\trans$ such that $\boldsymbol{u}\trans=\boldsymbol{y}\trans A$. Now $N_A\boldsymbol{v}=N_AA\boldsymbol{x}=0$ and $\boldsymbol{u}\trans N_A=\boldsymbol{y}\trans A N_A=0$. Likewise, $N_A\boldsymbol{v}=0$ and $\boldsymbol{u}\trans N_A=0$ imply that $\boldsymbol{v}$ and $\boldsymbol{u}\trans$ are in the image and coimage of $A$, respectively. 

There are a few useful properties of (co)vectors in the (co)image of $A$ and how they interact with $N_A$. Let $B$ be a non-singular matrix and $\boldsymbol{v}=A\boldsymbol{x}\in \text{im}\,A$ so $N_A\boldsymbol{v}=0$. One can then define $\boldsymbol{x'}=B^{-1}\boldsymbol{x}$ so $N_{AB}\boldsymbol{v} = N_{AB}AB\boldsymbol{x'} = 0$. Similarly, if $N_{AB}\boldsymbol{v} = 0$ so $\boldsymbol{v}=AB\boldsymbol{x'}\in\text{im}\,AB$, then $N_A\boldsymbol{v} = N_A AB\boldsymbol{x'}=0$. The analogous identity for the covector $\boldsymbol{u}\trans = \boldsymbol{y}\trans A\in \text{coim}\,A$ follows similarly. Thus,
\begin{subequations}
    \begin{align}
        N_{AB}\boldsymbol{v}=0 &\iff N_A\boldsymbol{v}=0\,, \\
        \boldsymbol{u}\trans N_{BA}=0 &\iff \boldsymbol{u}\trans N_A=0\,.
    \end{align}
\end{subequations}
Further, the $\implies$ direction holds when $B$ is singular.
Similarly for $N_{BA}\boldsymbol{v}=0$, there is some $\boldsymbol{x}$ such that $\boldsymbol{v}=BA\boldsymbol{x}$. Then, $N_AB^{-1}\boldsymbol{v} = N_A A\boldsymbol{x} = 0$. In the other direction, if $N_AB^{-1}\boldsymbol{v}=0$, then there is some $\boldsymbol{x}$ such that $A\boldsymbol{x}=B^{-1}\boldsymbol{v}$, and $N_{BA}\boldsymbol{v}=N_{AB}AB\boldsymbol{x}=0$. Again, the analogous identity for the covector $\boldsymbol{u}\trans$ follows similarly. Thus,
\begin{subequations}
    \begin{align}
        N_{BA}\boldsymbol{v} = 0 &\iff N_AB^{-1}\boldsymbol{v} = 0\,, \\
        \boldsymbol{u}\trans N_{AB} = 0 &\iff \boldsymbol{u}\trans B^{-1} N_A = 0\,.
    \end{align}
\end{subequations}

Under certain conditions, the Drazin inverse can adopt some useful properties from the matrix inverse. For a matrix $A$ and an invertible matrix $B$, let $\boldsymbol{v}\in\text{im}\left(A^\draz B\right)$ and $\boldsymbol{u}\trans\in\text{coim}\left(A^\draz B\right)$. One finds that 
\begin{equation}\label{eq:drazOfProd}
    \boldsymbol{u}\trans \left(A^\draz B\right)^\draz \boldsymbol{v} = \boldsymbol{u}\trans B^{-1} A \boldsymbol{v}\,,
\end{equation}
similar to the inverse of the product of invertible matrices. This identity can be proven by considering
\begin{align*}
    \left(A^\draz B\right)^\draz \left(A^\draz B\right) B^{-1}A &= \left(A^\draz B\right)^\draz \left(\ident - N_A \right) \\
    &= \left(\ident - N_{A^\draz B} \right) B^{-1}A\,.
\end{align*}
By equating the two right-hand equalities above,
\begin{align*}
    \boldsymbol{u}\trans \left(A^\draz B\right)^\draz \boldsymbol{v} - \boldsymbol{u}\trans \left(A^\draz B\right)^\draz N_A \boldsymbol{v} &= \boldsymbol{u}\trans B^{-1} A \boldsymbol{v} - \boldsymbol{u}\trans N_{A^\draz B} B^{-1} A \boldsymbol{v} \\
    \boldsymbol{u}\trans \left(A^\draz B\right)^\draz \boldsymbol{v} &= \boldsymbol{u}\trans B^{-1} A \boldsymbol{v}\,,
\end{align*}
using the fact that $\boldsymbol{u}\trans N_{A^\draz B}$ and $N_A \boldsymbol{v} = N_{A^\draz B} \boldsymbol{v} = 0$.

One may want to consider whether something similar to Eq.~\eqref{eq:drazOfProd} holds when $B$ is singular. For this, consider the special case in which $\ker B \subseteq \ker A$ and $\text{coker}\,B \subseteq \text{coker}\,A$; i.e., if one were to ignore the nullspace of $A$, then $B$ would appear non-singular as well. Thus, $N_AN_B = N_BN_A = N_B$ and $N_BA=AN_B=0$ because $N_B$ projects onto a more restrictive subspace. For $\boldsymbol{v}\in\text{im}\,A^{\draz}B\subseteq \text{im}\,A^{\draz} = \text{im}\,A^{\draz}(B+N_B)$ because $(B+N_B)$ is invertible. For $\boldsymbol{u}\trans\in \text{coim}\,A^{\draz}B$, there exists some $\boldsymbol{y}\trans$ such that
\begin{align*}
    \boldsymbol{u}\trans &= \boldsymbol{y}\trans A^{\draz}B \\
    &= \boldsymbol{y}\trans (A^{\draz}B + A^{\draz}N_A N_B) \\
    &= \boldsymbol{y}\trans A^{\draz}(B + N_B)\,,
\end{align*}
so $\boldsymbol{u}\trans\in\text{coim}\,A^{\draz}(B+N_B)$; further, following this calculation backwards shows that $\text{coim}\,A^{\draz}(B+N_B) = \text{coim}\,A^{\draz}B$ for $\ker B \subseteq \ker A$. Now, for $\boldsymbol{v}\in\text{im}\,A^{\draz}B$ and $\boldsymbol{u}\trans\in\text{coim}\,A^{\draz}B$, one can use Eq.~\eqref{eq:drazOfProd} with $B\to(B+N_B)$, which is invertible:
\begin{align*}
    \boldsymbol{u}\trans \left(A^\draz (B+N_B)\right)^\draz \boldsymbol{v} &= \boldsymbol{u}\trans (B+N_B)^{-1} A \boldsymbol{v} \\
    \boldsymbol{u}\trans \left(A^\draz B + A^\draz N_B \right)^\draz \boldsymbol{v} &= \boldsymbol{u}\trans \left(B^\draz A+N_BA\right) \boldsymbol{v}\,,
\end{align*}
but $A^{\draz}N_B = N_BA = 0$, thus
\begin{equation}
    \boldsymbol{u}\trans \left(A^\draz B\right)^\draz \boldsymbol{v} = \boldsymbol{u}\trans B^\draz A \boldsymbol{v}
\end{equation}
holds when the (co)kernel of $B$ is a subset of the (co)kernel of $A$, $\boldsymbol{v}\in\text{im}\,A^{\draz}B$, and $\boldsymbol{u}\trans\in\text{coim}\,A^{\draz}B$.

\section{Gillespie algorithm}\label{app:gillespie}

The Gillespie algorithm is a simple yet powerful tool to simulate continuous-time Markov processes~\cite{Gillespie_1976,Gillespie_1977} and has recently been extended to describe open quantum system dynamics~\cite{radaelli_gillespie_2023}. This algorithm is used to produce a set of random transitions and their respective times for a given process. 
This algorithm was used to generate the data shown in Figs.~\ref{fig:toy3_summ} and~\ref{fig:tho_summ}.

The time-continuous Markov model is characterized by the rate matrix $\rateMat$; wherein (for $x\neq y$) $\rateMat_{yx}$ is the rate at which state $x$ transitions to state $y$ and $-\rateMat_{xx}$ is the escape rate for state $x$ (i.e., the total rate at which the system transitions out of state $x$). Starting in state $x$, a step in the algorithm follows from generating two random numbers. First, from an exponential distribution characterized by $\rateMat_{xx}$, a time is generated for the transition. Second, a random state $y\neq x$ is selected with probability $\frac{\rateMat_{yx}}{-\rateMat_{xx}}$. Repeating this step many times can be used to simulate the Markov process.

The histograms shown in Fig.~\ref{fig:toy3_summ} and~\ref{fig:tho_summ} can be generated by integrating over the results of the simulated process. That is, $q_x$ is calculated by adding up the time intervals that start with a transition to state $x$ (normalized to the total time simulated).

\end{document}